\documentstyle[12pt,epsfig]{article} 

\newcommand{\agt}{\,\rlap{\lower 3.5 pt \hbox{$\mathchar \sim$}} \raise 1pt
 \hbox {$>$}\,}
\newcommand{\alt}{\,\rlap{\lower 3.5 pt \hbox{$\mathchar \sim$}} \raise 1pt
 \hbox {$<$}\,}
\newcommand{\re}{\mathop{\mathrm{Re}}\nolimits}

\catcode`@=11
\newcount\@tempcntc
\def\@citex[#1]#2{\if@filesw\immediate\write\@auxout{\string\citation{#2}}\fi
  \@tempcnta\z@\@tempcntb\m@ne\def\@citea{}\@cite{\@for\@citeb:=#2\do
    {\@ifundefined
       {b@\@citeb}{\@citeo\@tempcntb\m@ne\@citea\def\@citea{,}{\bf ?}\@warning
       {Citation `\@citeb' on page \thepage \space undefined}}%
    {\setbox\z@\hbox{\global\@tempcntc0\csname b@\@citeb\endcsname\relax}%
     \ifnum\@tempcntc=\z@ \@citeo\@tempcntb\m@ne
       \@citea\def\@citea{,}\hbox{\csname b@\@citeb\endcsname}%
     \else
      \advance\@tempcntb\@ne
      \ifnum\@tempcntb=\@tempcntc
      \else\advance\@tempcntb\m@ne\@citeo
      \@tempcnta\@tempcntc\@tempcntb\@tempcntc\fi\fi}}\@citeo}{#1}}
\def\@citeo{\ifnum\@tempcnta>\@tempcntb\else\@citea\def\@citea{,}%
  \ifnum\@tempcnta=\@tempcntb\the\@tempcnta\else
   {\advance\@tempcnta\@ne\ifnum\@tempcnta=\@tempcntb \else \def\@citea{--}\fi
    \advance\@tempcnta\m@ne\the\@tempcnta\@citea\the\@tempcntb}\fi\fi}
\catcode`@=12

\begin{document}

\title{\vskip-3cm{\baselineskip14pt
\centerline{\normalsize DESY 11--239\hfill ISSN 0418--9833}
\centerline{\normalsize December 2011\hfill}
}
\vskip1.5cm
Associated production of $Z$ and neutral Higgs bosons at the CERN Large Hadron
Collider}
\author{Bernd A. Kniehl$^1$ and Caesar P. Palisoc$^2$\\
{\normalsize $^1$ II. Institut f\"ur Theoretische Physik,
Universit\"at Hamburg,}\\
{\normalsize Luruper Chaussee 149, 22761 Hamburg, Germany}\\
{\normalsize $^2$ National Institute of Physics, University of the
Philippines,}\\
{\normalsize Diliman, Quezon City 1101, Philippines}}

\date{}

\maketitle

\thispagestyle{empty}

\begin{abstract}
We study the hadroproduction of a $CP$-even or $CP$-odd neutral Higgs boson in 
association with a $Z$ boson in the minimal supersymmetric extension of the
standard model (MSSM)
We include the contributions from quark-antiquark annihilation at the tree 
level and those from gluon-gluon fusion, which proceeds via quark and squark 
loops, and list compact analytic results.
We quantitatively analyze the hadronic cross sections at the CERN Large Hadron
Collider assuming a favorable supergravity-inspired MSSM scenario.

\medskip

\end{abstract}

\newpage

\section{Introduction}

The search for Higgs bosons is among the prime tasks of the CERN Large Hadron
Collider (LHC) \cite{Kunszt:1991qe}.
While the standard model (SM) of elementary-particle physics contains one
complex Higgs doublet, from which one neutral $CP$-even Higgs boson $H$
emerges in the physical particle spectrum after the spontaneous breakdown of
the electroweak symmetry, the Higgs sector of the minimal supersymmetric
extension of the SM (MSSM) consists of a two-Higgs-doublet model (2HDM) and
accommodates a quintet of physical Higgs bosons: the neutral $CP$-even $h^0$
and $H^0$ bosons, the neutral $CP$-odd $A^0$ boson, and the charged
$H^\pm$-boson pair.
At the tree level, the MSSM Higgs sector has two free parameters, which are
usually taken to be the mass $m_{A^0}$ of the $A^0$ boson and the ratio
$\tan\beta=v_2/v_1$ of the vacuum expectation values of the two Higgs
doublets.

In the following, we focus our attention on the $h^0$ ,$H^0$, and $A^0$
bosons, which we collectively denote by $\phi$.
A recent discussion of $H^\pm$-boson production at the LHC may be found in
Refs.~\cite{BarrientosBendezu:1998gd,BarrientosBendezu:1999vd,%
BarrientosBendezu:2000tu}
and the references cited therein.
The dominant source of $\phi$ bosons is their single production by $gg$
fusion, $gg\to\phi$, which is mediated by heavy-quark \cite{Georgi:1977gs} and
squark \cite{Dawson:1996xz} loops.
Another, less important mechanism of single $\phi$-boson production is
$b\bar b\to\phi$ \cite{Dicus:1988cx}.
The $\phi$ bosons thus produced have essentially zero transverse momentum
($p_T$).
In order for the $\phi$ bosons to obtain finite $p_T$, they need to be 
produced in association with one or more other particles or jets ($j$).
In leading order (LO), $j\phi$ associated production proceeds through the
partonic subprocesses $gg\to g\phi$, $gq\to q\phi$, and $q\bar q\to g\phi$,
which again involve heavy-quark \cite{Ellis:1987xu,Brein:2003df} and squark
\cite{Brein:2003df,Muhlleitner:2006wx} loops.
Alternatively, the $\phi$ bosons can be produced, with interesting rates, in
association with
(i) a dijet via intermediate-boson fusion,
$qq^\prime\to qq^\prime V^*V^*\to qq^\prime\phi$, where $q$ and $q^\prime$
stand for any light flavor of quark or antiquark, $V=W^\pm,Z$, and virtual
particles are marked by an asterisk \cite{Cahn:1983ip};
(ii) a quark-antiquark pair of heavy flavor $Q=t,b$ via
$gg,q\bar q\to Q\bar Q\phi$ \cite{Kunszt:1984ri};
(iii) an intermediate boson via
$q\bar q^\prime\to W^\pm\phi$
\cite{Glashow:1978ab,Han:1991ia,Ohnemus:1992bd,Djouadi:1999ht},
$q\bar q\to Z\phi$
\cite{Glashow:1978ab,Han:1991ia,Djouadi:1999ht,Kniehl:1990iva,Yin:2002sq,%
Yang:2003kr,Kao:2004vp,Li:2005qna},
and $gg\to Z\phi$
\cite{Kniehl:1990iva,Yin:2002sq,Yang:2003kr,Dicus:1988yh,Kao:1991xg,%
Kao:2003jw,Brein:2003wg};
or (iv) another, possibly different $\phi$ boson via
$q\bar q\to\phi_1\phi_2$
\cite{Djouadi:1999ht,Dawson:1998py,BarrientosBendezu:2001di} and
$gg\to\phi_1\phi_2$
\cite{BarrientosBendezu:2001di,Plehn:1996wb,Belyaev:1999mx}.
Note that, due to the absence of $A^0VV$ couplings at the tree level,
$qq^\prime\to qq^\prime V^*V^*\to qq^\prime\phi$ and the Drell-Yan processes
$q\bar q^\prime\to W^{\pm*}\to W^\pm\phi$ and $q\bar q\to Z^*\to Z\phi$ are
not possible for $\phi=A^0$.
The partonic suprocesses $gg\to Z\phi$ and $gg\to\phi_1\phi_2$ are mediated
by heavy-quark
\cite{Kniehl:1990iva,Yin:2002sq,Yang:2003kr,Dicus:1988yh,Kao:1991xg,%
Kao:2003jw,Brein:2003wg,BarrientosBendezu:2001di,Plehn:1996wb,Belyaev:1999mx}
and squark loops \cite{Yin:2002sq,BarrientosBendezu:2001di,Belyaev:1999mx}.
Comprehensive reviews of quantum corrections to Higgs-boson production within
the SM and MSSM may be found in Refs.~\cite{Kniehl:1993ay,Spira:1997dg},
respectively.

In this paper, we revisit $Z\phi$ associated hadroproduction via gluon fusion
in the MSSM.
In the SM case, mutual agreement between three independent calculations
\cite{Kniehl:1990iva,Dicus:1988yh,Brein:2003wg} has been established, and
compact formulae for the partonic cross section are available
\cite{Kniehl:1990iva}.
In the MSSM, the status is much less advanced and, perhaps, somewhat
unsatisfactory.
As for $gg\to Z\phi$ with $\phi=h^0,H^0$, there exists only one analysis so
far \cite{Yang:2003kr}, which has not yet been verified by other authors.
The analytic expressions presented in Ref.~\cite{Yang:2003kr} are rather
complicated; they involve 14 form factors.
In order to translate the SM results
\cite{Kniehl:1990iva,Dicus:1988yh,Brein:2003wg} to the case of $gg\to Z\phi$
with $\phi=h^0,H^0$ in the MSSM, it is not sufficient to adjust the $HZZ$ and
$Hqq$ couplings; it is also necessary to include certain quark triangle
diagrams with an $A^0$ boson in the $s$ channel [see Fig.~\ref{fig:tree}(a)].
On the other hand, the squark loop contributions to these two MSSM processes
vanish \cite{Yang:2003kr} for reasons explained below.
As for $gg\to ZA^0$, the quark loop contributions were first studied on the
basis of a numerical evaluation \cite{Kao:1991xg}, which was recently employed
for phenomenological signal-versus-background analyses taking into account the
subsequent $Z\to l^+l^-$ and $A^0\to b\bar b$ decays
\cite{Kao:2004vp,Kao:2003jw}.
An independent analysis, including also the squark loops, was reported in
Ref.~\cite{Yin:2002sq}, which does not contain an analytic expression for the
partonic cross section either.
Unfortunately, comparisons between
Refs.~\cite{Yin:2002sq,Kao:2004vp,Kao:1991xg,Kao:2003jw} are not discussed in
these papers.
Recently, the helicity amplitudes of $gg\to Z\phi$ with $\phi=h^0,H^0,A^0$
were analyzed in Ref.~\cite{Gounaris:2009cc} with regard to the asymptotic
helicity conservation property of supersymmetry, and their real and imaginary
parts were graphically presented at a specific scattering angle as functions
of the center-of-mass (c.m.) energy.

With the Higgs hunt at the LHC being in full swing, it is an urgent matter to
consolidate our knowledge of $Z\phi$ associated hadroproduction via gluon
fusion in the MSSM, which is the motivation of this paper.
Specifically, we present compact analytic expressions, also for the partonic
cross sections of $q\bar q,b\bar b\to Z\phi$
\cite{Yin:2002sq,Yang:2003kr,Kao:2004vp,Li:2005qna}, and perform a detailed
numerical analysis using up-to-date input.

The importance of $b\bar b$-initiated subprocesses for Higgs-boson production
has been variously emphasized in the literature, in particular, in connection 
with the final states $\phi$ \cite{Dicus:1988cx}, $\phi_1\phi_2$
\cite{BarrientosBendezu:2001di}, $H^+H^-$ \cite{BarrientosBendezu:1999gp}, and
$W^\pm H^\mp$ \cite{BarrientosBendezu:1998gd,Dicus:1989vf}.
These subprocesses receive contributions from Feynman diagrams involving
$b$-quark Yukawa couplings, which are generally strong for large values of
$\tan\beta$.
(The $\bar btH^-$ and $\bar tbH^+$ couplings are also strong for small values 
of $\tan\beta$.)
If the two final-state particles couple to a $Z$ boson (or photon), as is the
case for the final states $h^0A^0$, $H^0A^0$, $Zh^0$, $ZH^0$, and $H^+H^-$,
then there are additional contributions from Drell-Yan-type diagrams, which
are already present for the light flavors $q=u,d,s,c$.
However, diagrams of the latter type are absent for the final states $h^0h^0$,
$h^0H^0$, $H^0H^0$, $A^0A^0$, $ZA^0$, and $W^\pm H^\mp$, which can still be
produced through $b\bar b$ annihilation.

As for $b\bar b$ annihilation, it should be noted that the treatment of bottom
as an active flavor inside the colliding hadrons leads to an effective
description, which comprises contributions from the higher-order subprocesses
$gb\to Z\phi b$, $g\bar b\to Z\phi\bar b$, and $gg\to Z\phi b\bar b$.
If all these subprocesses are to be explicitly included along with
$b\bar b\to Z\phi$, then it is necessary to employ a judiciously
subtracted $b$-quark PDF in order to avoid double counting
\cite{Dicus:1988cx,Gunion:1986pe}.
The evaluation of $b\bar b\to Z\phi$ with an unsubtracted $b$-quark PDF
is expected to slightly overestimate the true cross section
\cite{Dicus:1988cx,Gunion:1986pe}.
For simplicity, we shall nevertheless adopt this effective approach in our
analysis, keeping in mind that a QCD-correction factor below unity is to be 
applied.
In fact, such a behavior has recently been observed for $b\bar b\to ZA^0$
\cite{Li:2005qna}.

In order to reduce the number of unknown supersymmetric input parameters, we 
adopt a scenario where the MSSM is embedded in a grand unified theory (GUT)
involving supergravity (SUGRA) \cite{Djouadi:1996pj}.
The MSSM thus constrained is characterized by the following parameters at the
GUT scale, which come in addition to $\tan\beta$ and $m_{A^0}$: the universal
scalar mass $m_0$, the universal gaugino mass $m_{1/2}$, the trilinear
Higgs-sfermion coupling $A$, the bilinear Higgs coupling $B$, and the
Higgs-higgsino mass parameter $\mu$.
Notice that $m_{A^0}$ is then not an independent parameter anymore, but it is
fixed through the renormalization group equation.
The number of parameters can be further reduced by making additional
assumptions.
Unification of the $\tau$-lepton and $b$-quark Yukawa couplings at the GUT
scale leads to a correlation between $m_t$ and $\tan\beta$.
Furthermore, if the electroweak symmetry is broken radiatively, then $B$ and
$\mu$ are determined up to the sign of $\mu$.
Finally, it turns out that the MSSM parameters are nearly independent of the
value of $A$, as long as $|A|\alt500$~GeV at the GUT scale.

This paper is organized as follows.
In Sec.~\ref{sec:two}, we list the LO cross sections of $q\bar q\to Z\phi$, 
including the Yukawa-enhanced contributions for $q=b$, and those of
$gg\to Z\phi$, including both quark and squark loop contributions, in the
MSSM.
The relevant quark and squark loop form factors are relegated to
Appendix~\ref{sec:b}.
In Sec.~\ref{sec:three}, we present quantitative predictions for the inclusive
cross sections of $pp\to Z\phi+X$ at the LHC adopting a favorable
SUGRA-inspired MSSM scenario.
Sec.~\ref{sec:four} contains our conclusions.
For the reader's convenience, the relevant Feynman rules are summarized in
Appendix~\ref{sec:a}.

\section{\label{sec:two}Analytic Results}

In this section, we present the LO cross sections of the partonic
subprocesses $q\bar q\to Z\phi$ and $gg\to Z\phi$, where $\phi=h^0,H^0,A^0$,
in the MSSM.
We work in the parton model of QCD with $n_f=5$ active quark flavors
$q=u,d,s,c,b$, which we take to be massless.
However, we retain the $b$-quark Yukawa couplings at their finite values, in 
order not to suppress possibly sizable contributions.
We adopt the MSSM Feynman rules from Ref.~\cite{Haber:1984rc}.
The couplings of the $Z$ and $\phi$ bosons to quarks,
$v_{Zqq}$, $a_{Zqq}$, and $g_{\phi qq}$, are given in Eq.~(5) of 
Ref.~\cite{BarrientosBendezu:1999gp} and Eq.~(A3) of
Ref.~\cite{BarrientosBendezu:2001di}, respectively.
As for the $\phi ZZ$ couplings, $g_{h^0ZZ}$ and $g_{H^0ZZ}$ are given by
Eq.~(\ref{eq:hzz}) in Appendix~\ref{sec:a}, while the $A^0ZZ$ coupling
vanishes at tree level.
The $h^0A^0Z$ and $H^0A^0Z$ couplings, $g_{h^0A^0Z}$ and $g_{H^0A^0Z}$, may be
found in Eq.~(A2) of Ref.~\cite{BarrientosBendezu:2001di}.
For each quark flavor $q$ there is a corresponding squark flavor $\tilde q$,
which comes in two mass eigenstates $i=1,2$.
The masses $m_{\tilde q_i}$ of the squarks and their trilinear couplings to
the $\phi$ bosons, $g_{\phi\tilde q_i\tilde q_j}$, are listed
in Eqs.~(A.5), (A.7), and (A.8) and in Table~1 of
Ref.~\cite{Hempfling:1993ru}\footnote{%
In Ref.~\cite{Hempfling:1993ru}, $m_{\tilde q_i}$ and
$g_{\phi\tilde q_i\tilde q_j}$ are called $M_{\tilde Qa}$ and
$\tilde V_{Qab}^\phi/g$, respectively.}, Eq.~(A.2)
of Ref.~\cite{BarrientosBendezu:1999gp}, and Eq.~(A4) of
Ref.~\cite{BarrientosBendezu:2001di}, respectively.

Considering the generic partonic subprocess $ab\to Z\phi$, we denote the
four-momenta of the incoming partons, $a$ and $b$, and the outgoing $Z$ and 
$\phi$ bosons by $p_a$, $p_b$, $p_Z$, and $p_\phi$, respectively, and define
the partonic Mandelstam variables as $s=(p_a+p_b)^2$, $t=(p_a-p_Z)^2$, and
$u=(p_b-p_Z)^2$.
The on-shell conditions read $p_a^2=p_b^2=0$, $p_Z^2=m_Z^2=z$, and 
$p_\phi^2=m_\phi^2=h$.
Four-momentum conservation implies that $s+t+u=z+h$.
Furthermore, we have $sp_T^2=tu-zh=N$, where $p_T$ is the absolute value of
transverse momentum common to the $Z$ and $\phi$ bosons in the c.m.\ frame.

\begin{figure}[ht]
\begin{center}
\begin{tabular}{ccc}
\parbox{4cm}{\epsfig{figure=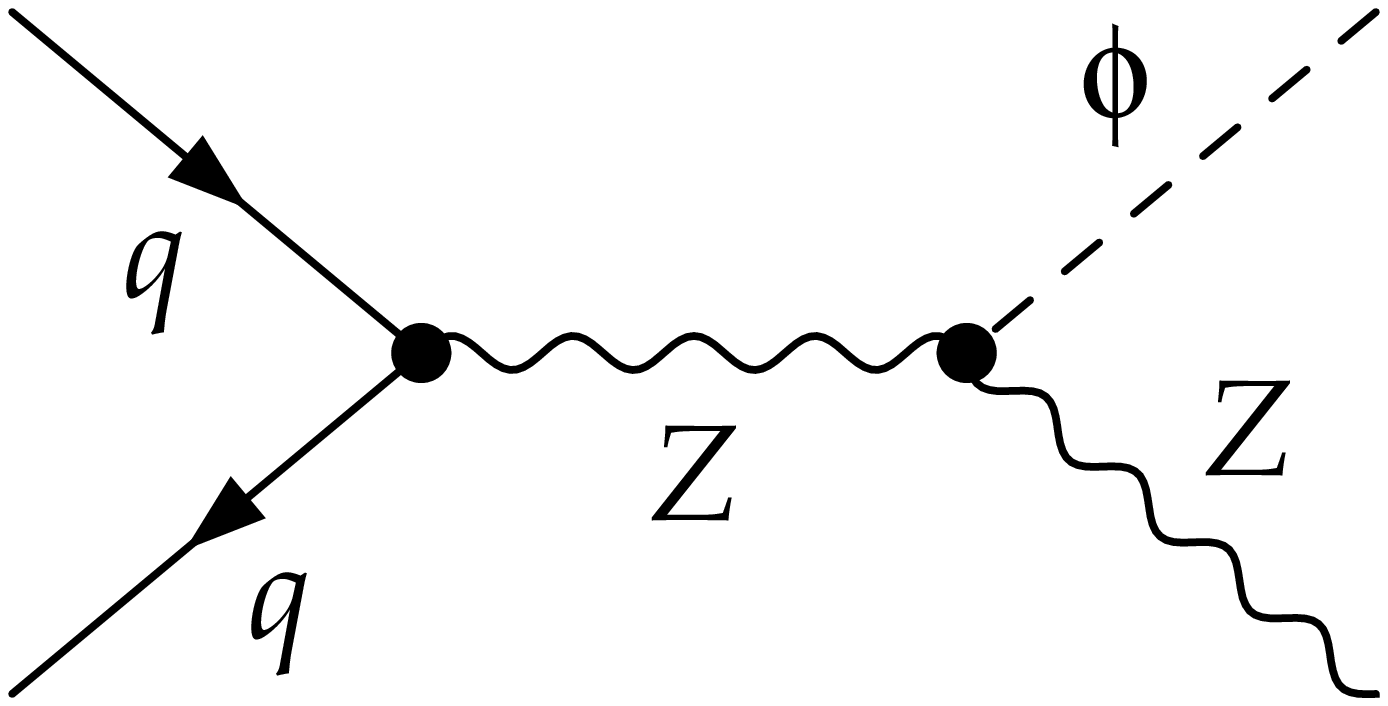,width=4cm}} &
\parbox{4cm}{\epsfig{figure=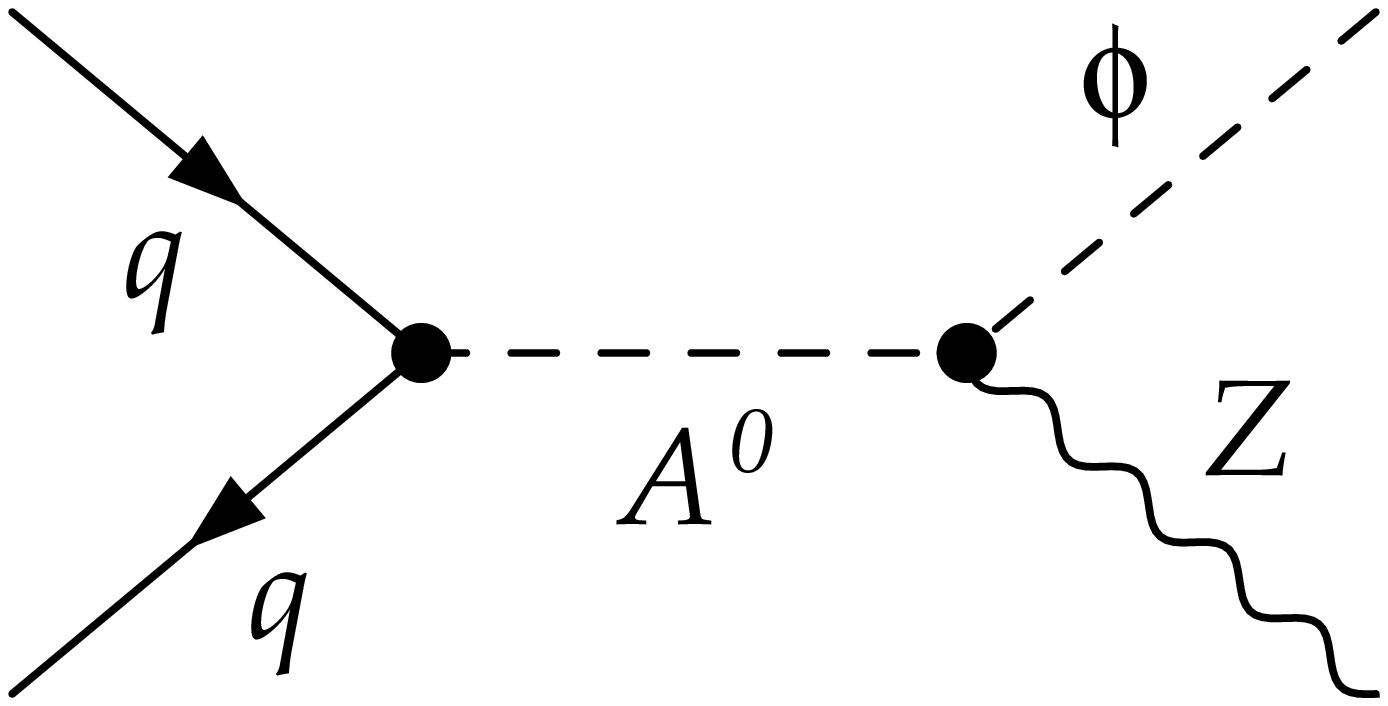,width=4cm}} &
\parbox{4cm}{\epsfig{figure=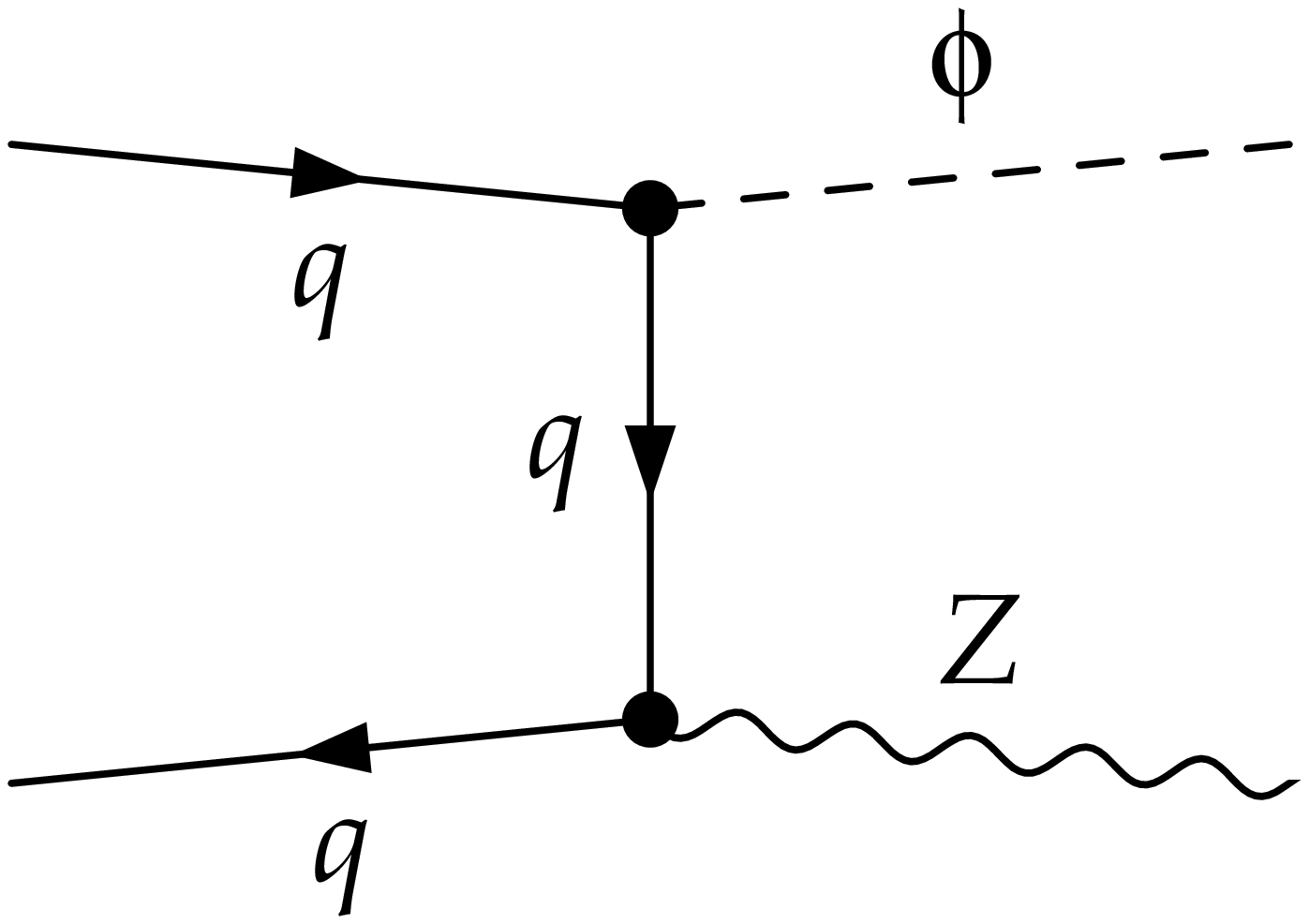,width=4cm}} \\
 & (a) & \\
\end{tabular}
\begin{tabular}{ccc}
\parbox{4cm}{\epsfig{figure=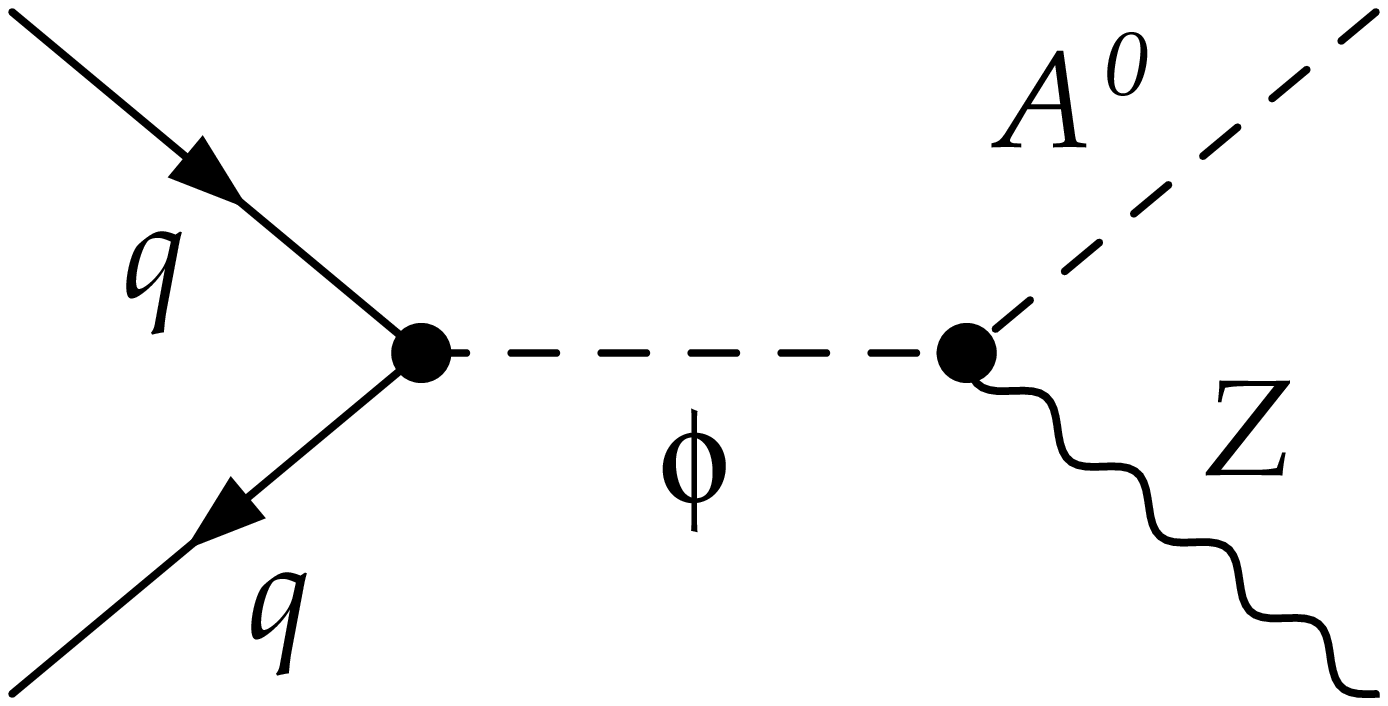,width=4cm}} & &
\parbox{4cm}{\epsfig{figure=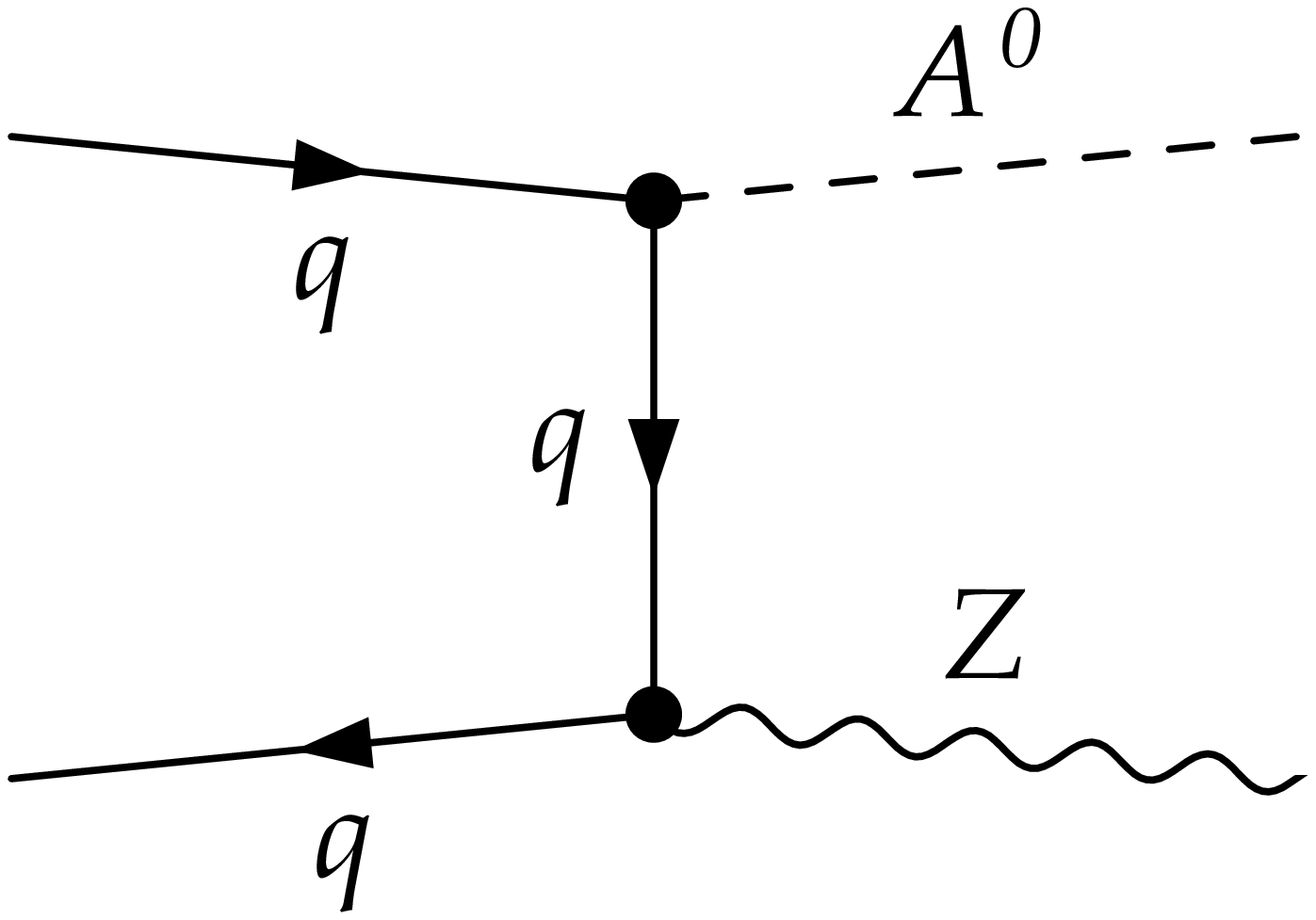,width=4cm}} \\
& (b) &
\end{tabular}
\caption{Tree-level Feynman diagrams for $q\bar q\to Z\phi$, with (a)
$\phi=h^0,H^0$ and (b) $\phi=A^0$, in the MSSM.}
\label{fig:tree}
\end{center}
\end{figure}

The tree-level diagrams for $b\bar b\to Z\phi$ with $\phi=h^0,H^0$ and
$\phi=A^0$ are depicted in Fig.~\ref{fig:tree}(a) and (b), respectively.
As already mentioned above, there is no Drell-Yan diagram in 
Fig.~\ref{fig:tree}(b) because of the absence of a $A^0ZZ$ coupling at the
tree level.
The differential cross sections for the first class of partonic subprocesses
may be generically written as
\begin{eqnarray}
\frac{d\sigma}{dt}\left(b\bar b\to Z\phi\right)&=&\frac{G_F^2c_w^4z}{3\pi s}
\left[\left(2z+p_T^2\right)g_{\phi ZZ}^2\left(v_{Zbb}^2+a_{Zbb}^2\right)
|{\cal P}_Z(s)|^2+\lambda|P|^2
\vphantom{\frac{1}{t}}\right.
\nonumber\\
&&{}-\left.4sp_T^2\left(\frac{1}{t}+\frac{1}{u}\right)g_{\phi bb}a_{Zbb}
\re P
+g_{\phi bb}^2\left(v_{Zbb}^2T_++a_{Zbb}^2T_-\right)\right],
\label{eq:bbzh}
\end{eqnarray}
where $G_F$ is Fermi's constant, $c_w=m_W/m_Z$ is the cosine of the weak mixing
angle, $\lambda=s^2+z^2+h^2-2(sz+zh+hs)$, and
\begin{eqnarray}
P&=&g_{\phi A^0Z}g_{A^0bb}{\cal P}_{A^0}(s),
\nonumber\\
T_\pm&=&2\pm2+2p_T^2\left[z\left(\frac{1}{t}\pm\frac{1}{u}\right)
\mp\frac{2s}{tu}\right].
\end{eqnarray}
Here,
\begin{equation}
{\cal P}_{X}(s)=\frac{1}{s-m_X^2+im_X\Gamma_X}
\end{equation}
is the propagator function of particle $X$, with mass $m_X$ and total decay
width $\Gamma_X$.
For the second class of partonic subprocesses, we have
\begin{eqnarray}
\frac{d\sigma}{dt}\left(b\bar b\to ZA^0\right)&=&\frac{G_F^2c_w^4z}{3\pi s}
\left[\lambda|S|^2
-4sp_T^2\left(\frac{1}{t}+\frac{1}{u}\right)g_{A^0bb}a_{Zbb}\re S
\right.
\nonumber\\
&&{}+\left.\vphantom{\frac{1}{t}}
g_{A^0bb}^2\left(v_{Zbb}^2T_++a_{Zbb}^2T_-\right)\right],
\label{eq:bbza}
\end{eqnarray}
where
\begin{equation}
S=g_{h^0A^0Z}g_{h^0bb}{\cal P}_{h^0}(s)+g_{H^0A^0Z}g_{H^0bb}{\cal P}_{H^0}(s).
\end{equation}
As for $Zh^0$ and $ZH^0$ production, there are also sizable contributions
from $q\bar q$ annihilation via a virtual $Z$ boson for the quarks of the
first and second generations, $q=u,d,s,c$, whose Yukawa couplings are
negligibly small.
The corresponding Drell-Yan cross sections are obtained from
Eq.~(\ref{eq:bbzh}) by putting $P=T_\pm=0$ and substituting $b\to q$.
The resulting expression agrees with Eq.~(2.8) of Ref.~\cite{Kniehl:1990iva},
appropriate for $q\bar q\to ZH$ in the SM, after adjusting the $HZZ$ coupling.
The full tree-level cross sections are then obtained by complementing the
$b\bar b$-initiated cross sections of Eq.~(\ref{eq:bbzh}) with the Drell-Yan
cross sections for $q=u,d,s,c$.

\begin{figure}[ht]
\begin{center}
\begin{tabular}{ccc}
\parbox{4cm}{\epsfig{figure=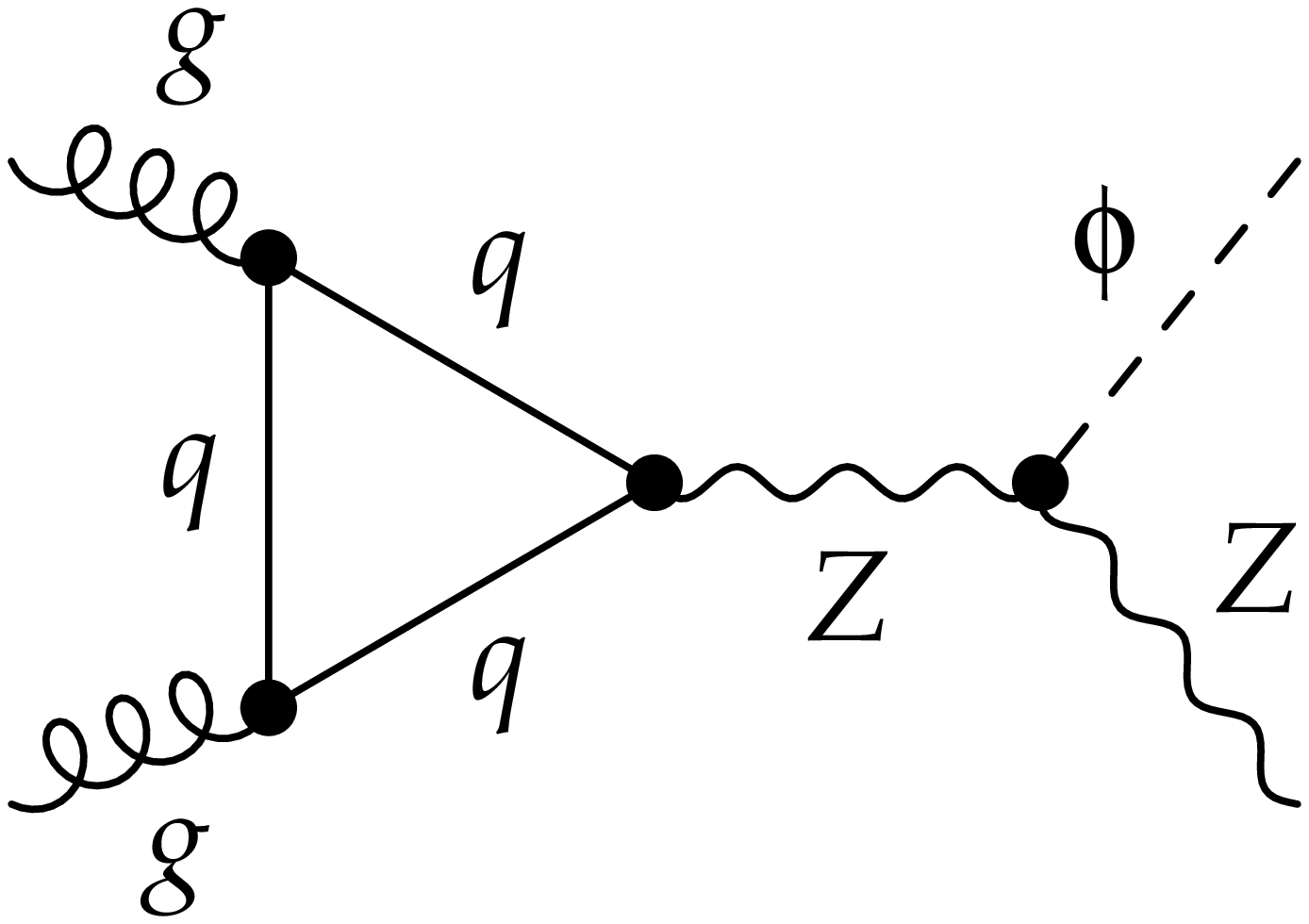,width=4cm}} &
\parbox{4cm}{\epsfig{figure=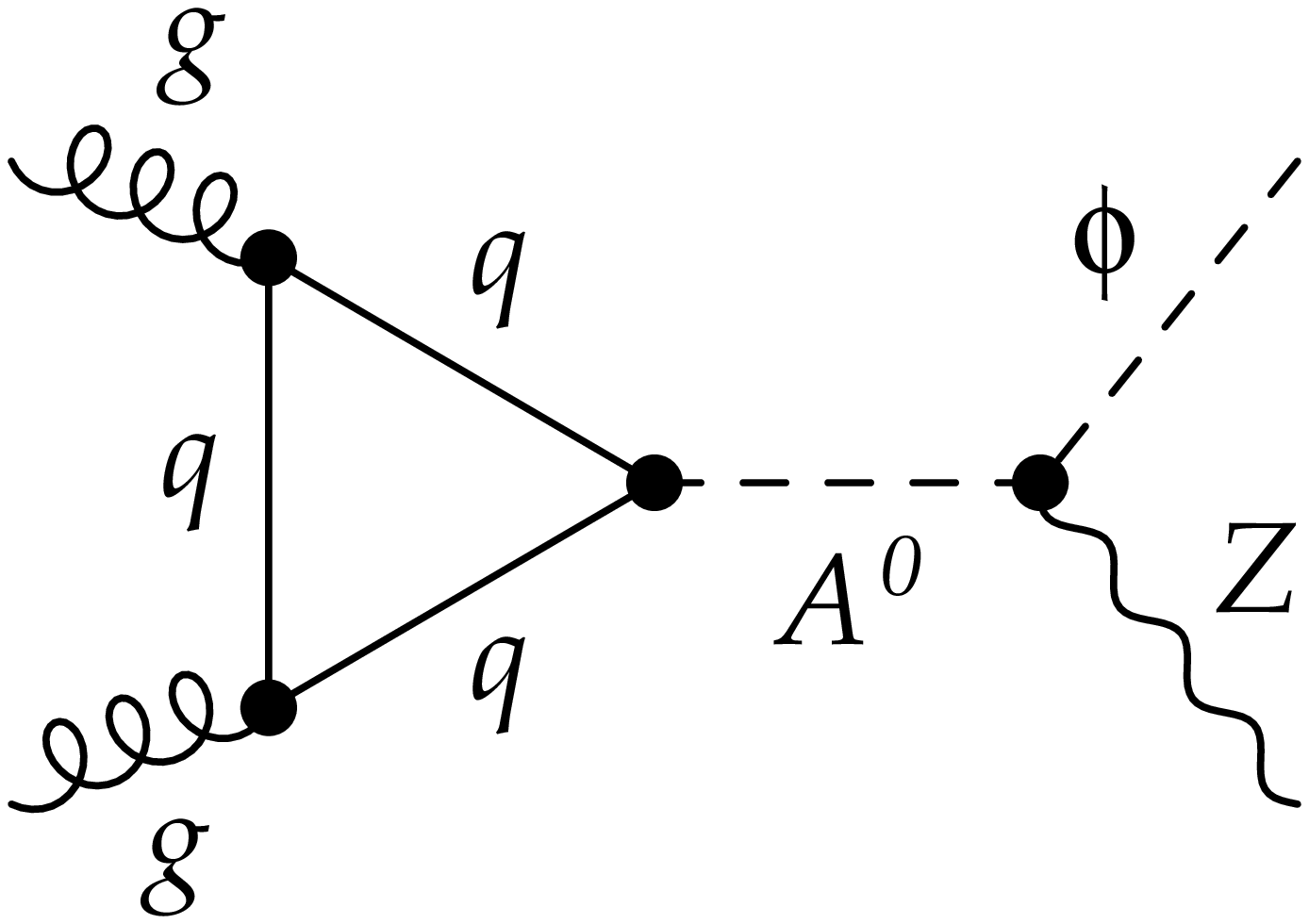,width=4cm}} &
\parbox{4cm}{\epsfig{figure=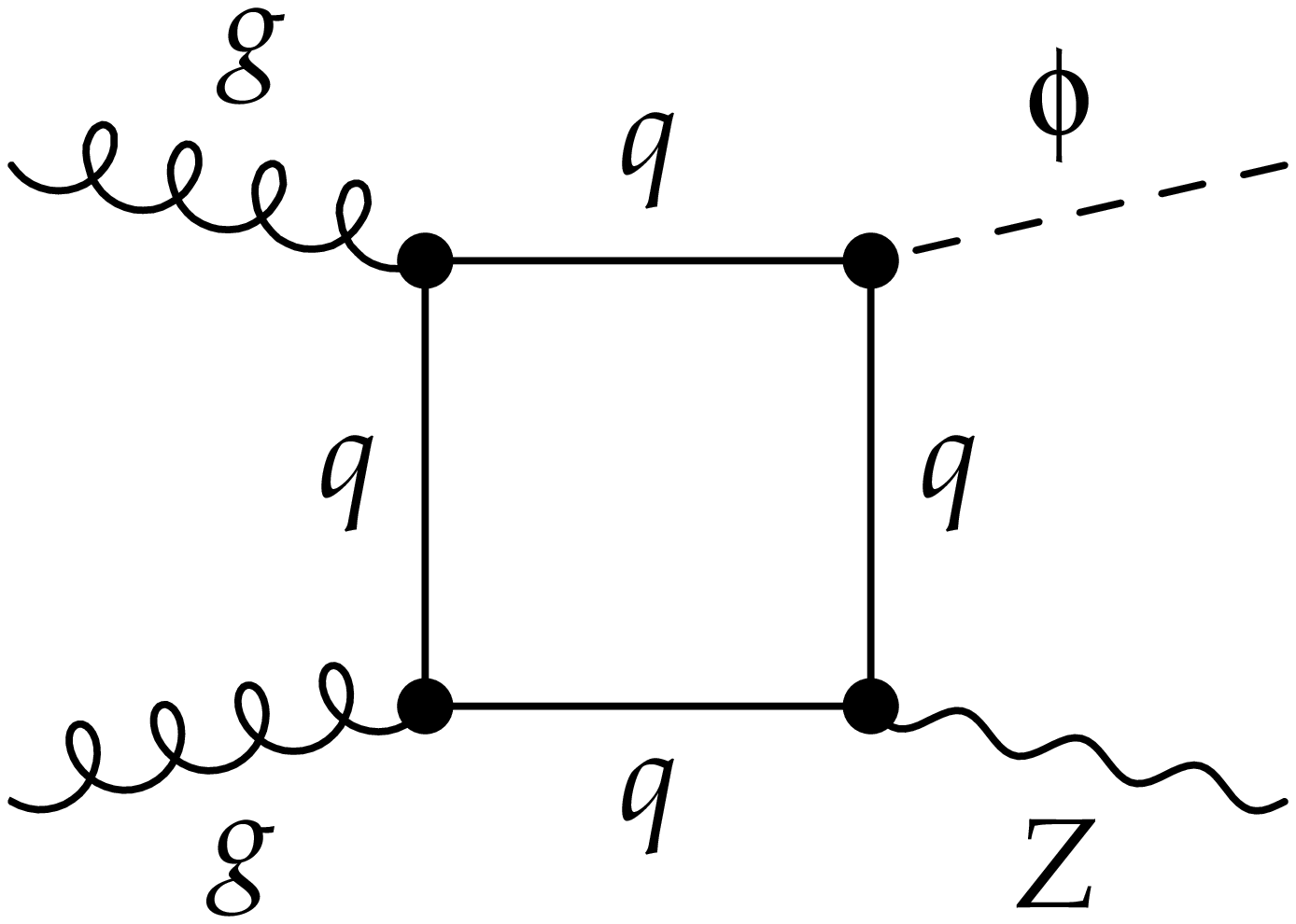,width=4cm}} \\
 & (a) &
\end{tabular}
\begin{tabular}{ccc}
\parbox{4cm}{\epsfig{figure=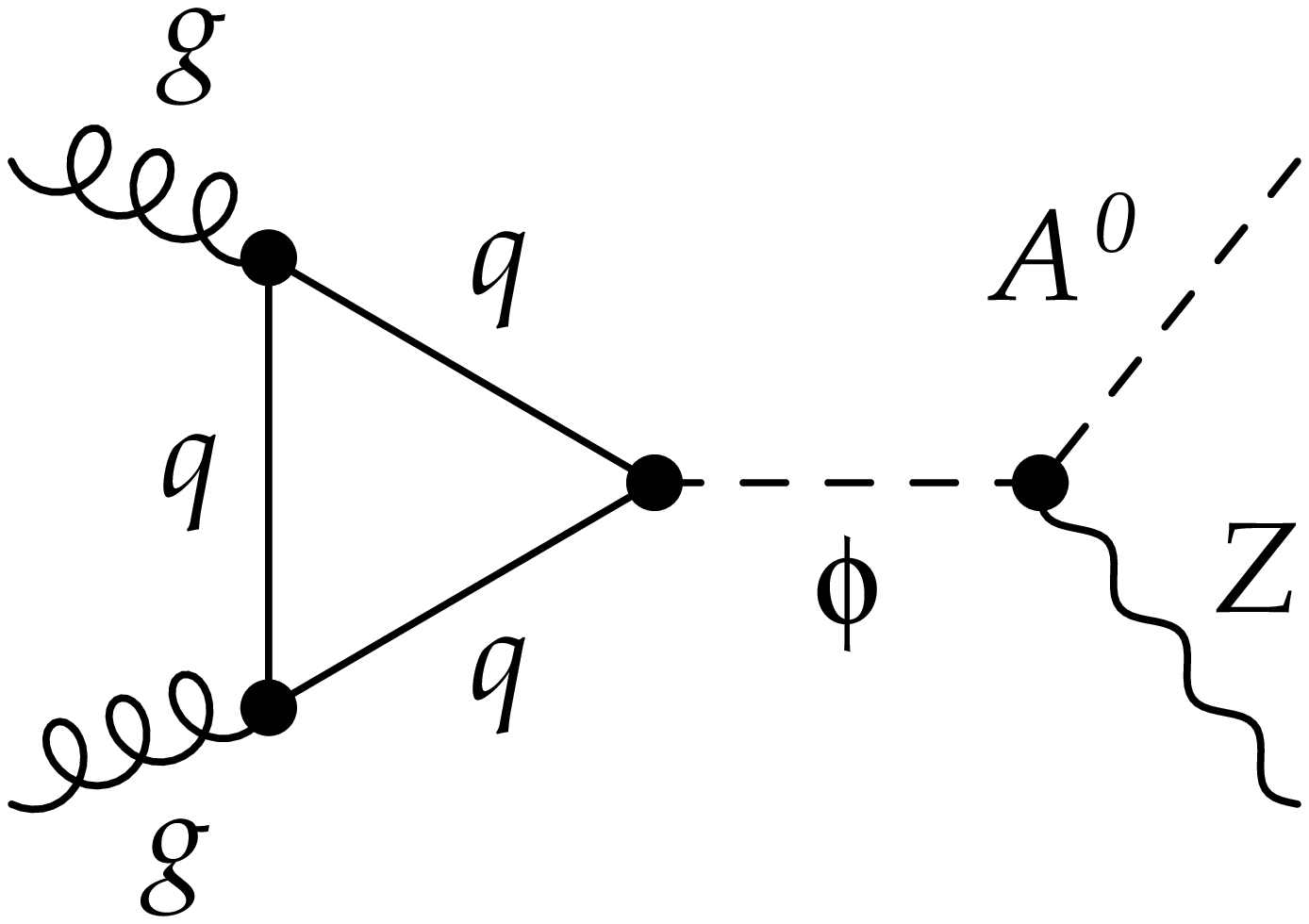,width=4cm}} & &
\parbox{4cm}{\epsfig{figure=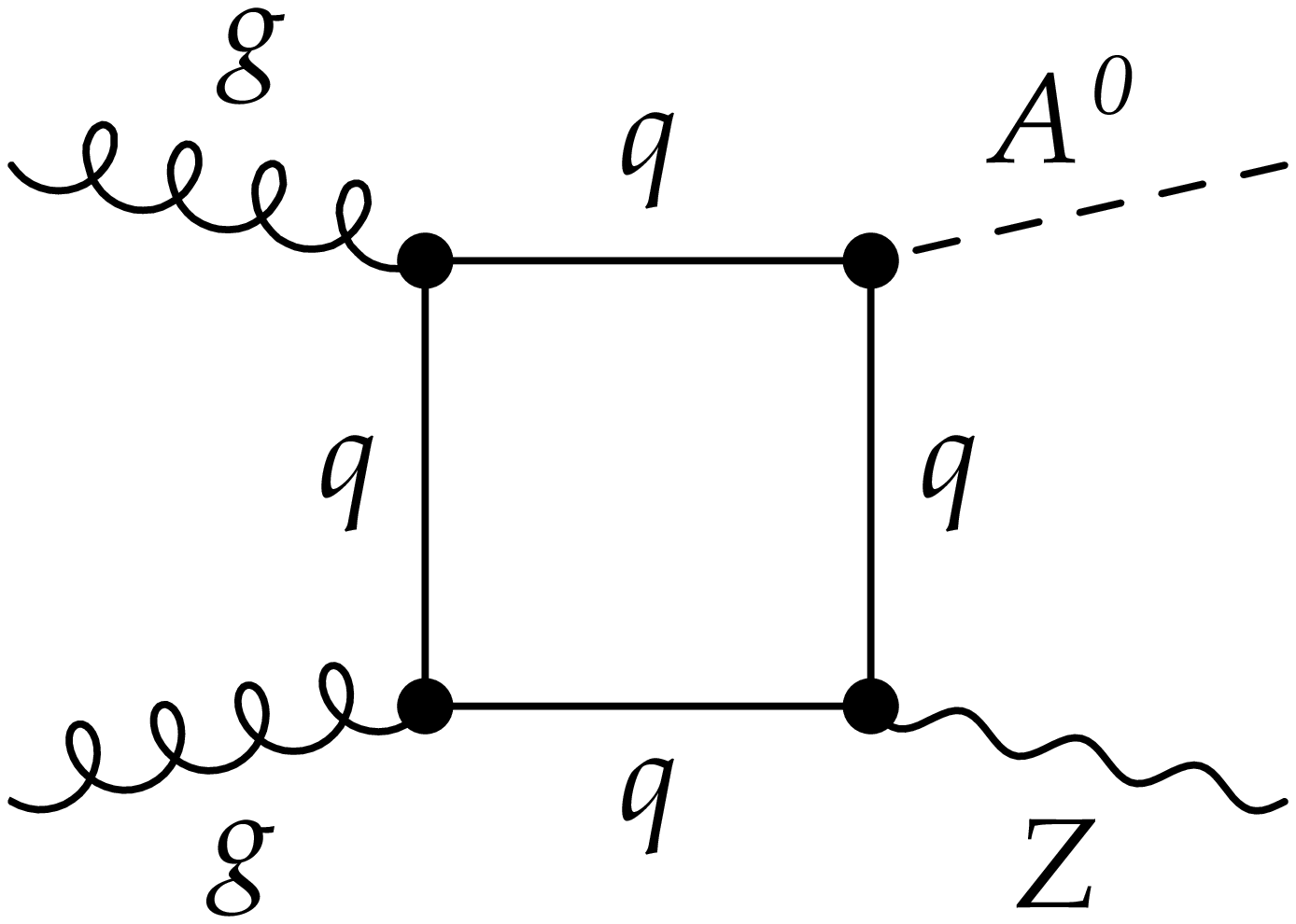,width=4cm}} \\
\parbox{4cm}{\epsfig{figure=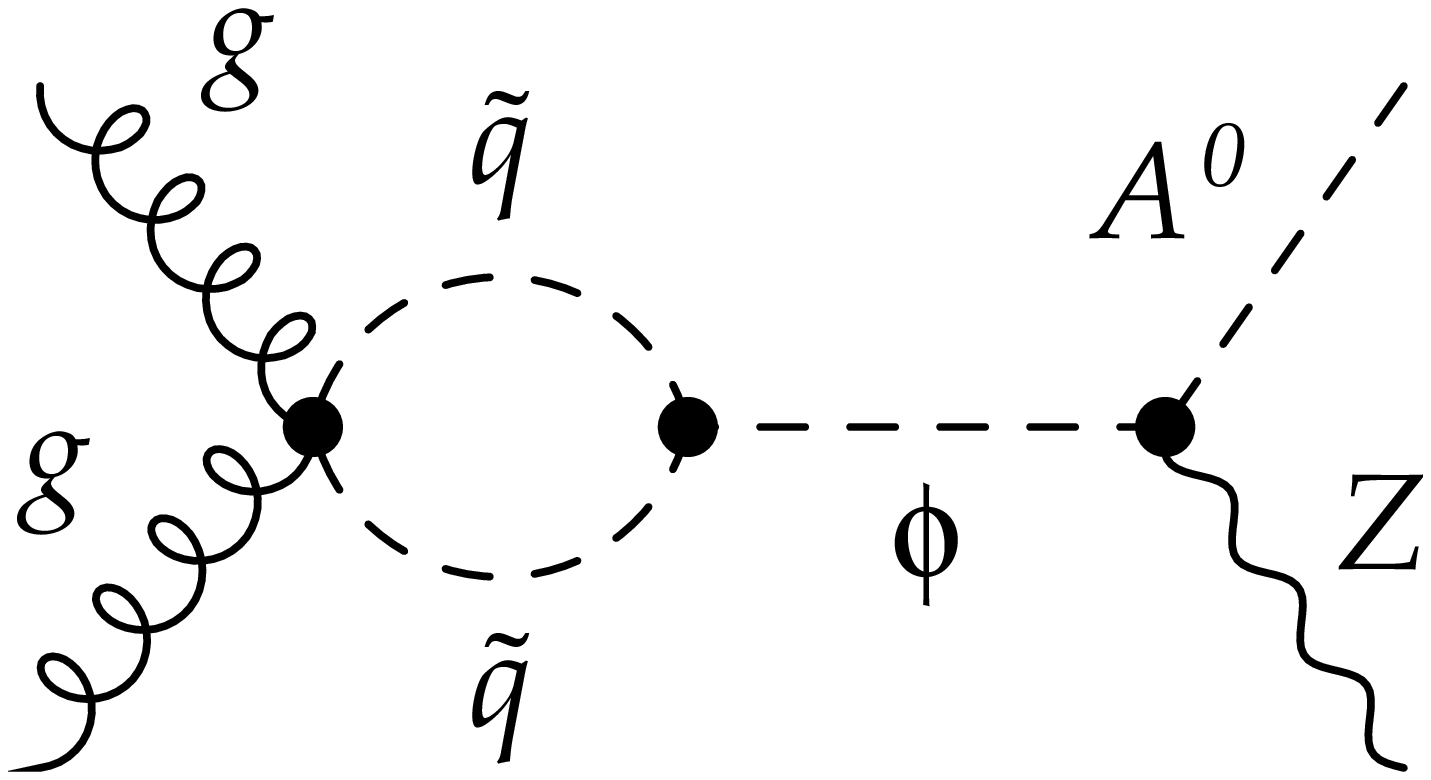,width=4cm}} & &
\parbox{4cm}{\epsfig{figure=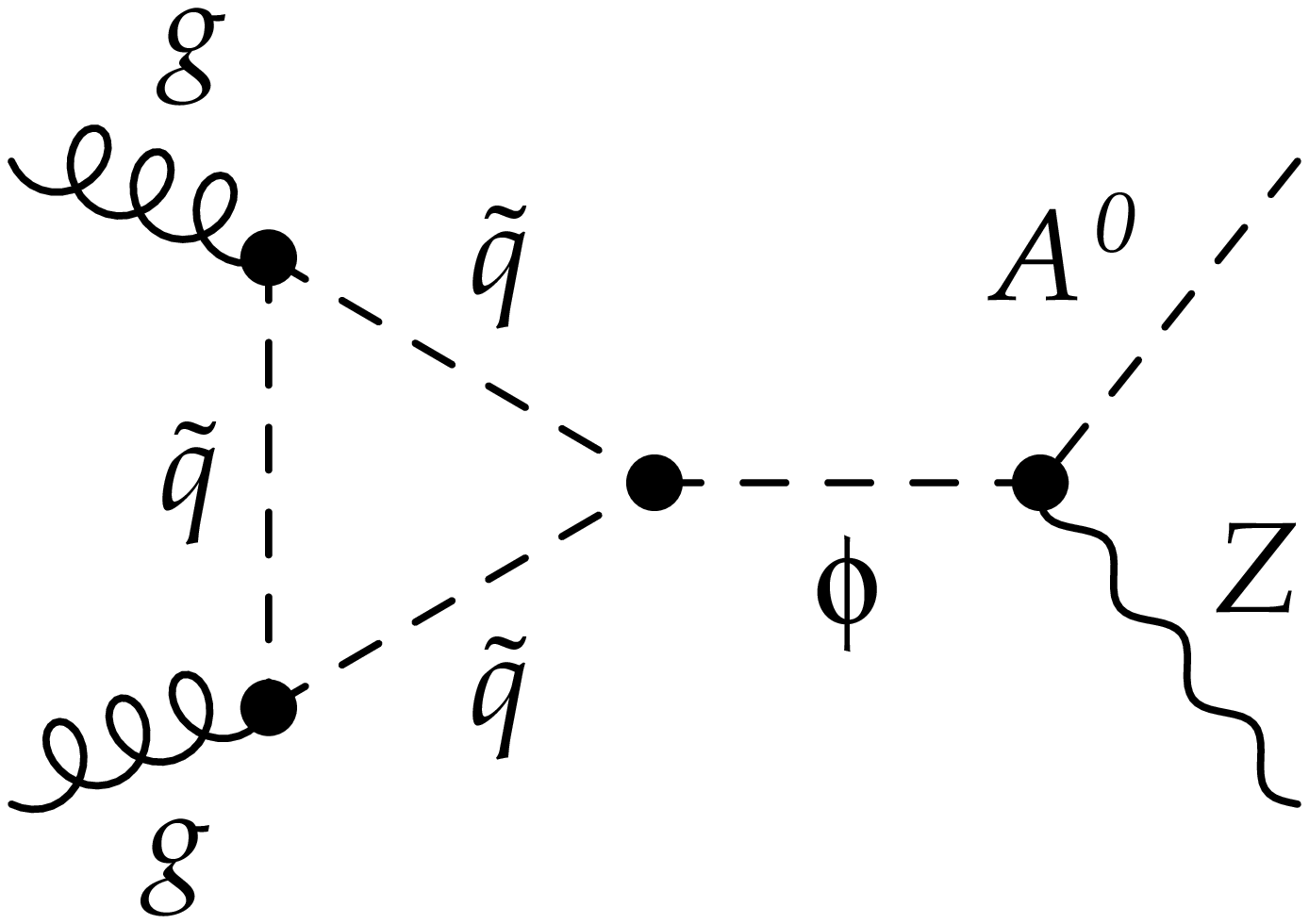,width=4cm}} \\
 & (b) &
\end{tabular}
\caption{One-loop Feynman diagrams for $gg\to Z\phi$, with (a) $\phi=h^0,H^0$
and (b) $\phi=A^0$, due to virtual quarks and squarks in the MSSM.}
\label{fig:loop}
\end{center}
\end{figure}

The non-vanishing one-loop diagrams pertinent to $gg\to Z\phi$, with
$\phi=h^0,H^0$ and $\phi=A^0$ are depicted in Figs.~\ref{fig:loop}(a) and (b),
respectively.
As already mentioned in the Introduction, the presence of the quark triangle
diagrams involving an $s$-channel $A^0$-boson exchange in
Fig.~\ref{fig:loop}(a) represents a qualitatively new feature of the MSSM as
compared to the SM.
Furthermore, similarly to Fig.~\ref{fig:tree}(b), quark triangle diagrams with
an $s$-channel $Z$-boson exchange do not appear in Fig.~\ref{fig:loop}(b).
In the following, we refer to a squark loop diagram involving an $s$-channel
propagator as a triangle diagram.
The residual squark loop diagrams are regarded to be of box type.
The squark triangle and box diagrams for $gg\to Z\phi$ with $\phi=h^0,H^0$
vanish, and so do the squark box diagrams for $gg\to ZA^0$.
This may be understood as follows.
(i) The $g_{g\tilde q_i\tilde q_j}$, $g_{gg\tilde q_i\tilde q_j}$,
$g_{gZ\tilde q_i\tilde q_j}$, and $g_{Z\tilde q_i\tilde q_j}$ couplings are
symmetric in $i$ and $j$, while the $g_{A^0\tilde q_i\tilde q_j}$ coupling is
antisymmetric \cite{Rosiek:1989rs}.
Thus, squark loops connecting gluons and $Z$ bosons with an odd number of $A^0$
bosons vanish upon summation over $i$ and $j$.
(ii) The $g_{g\tilde q_i\tilde q_j}$ and $g_{Z\tilde q_i\tilde q_j}$ couplings
are linear in the squark four-momenta, while the $g_{gg\tilde q_i\tilde q_j}$,
$g_{gZ\tilde q_i\tilde q_j}$, and $g_{\phi\tilde q_i\tilde q_j}$ couplings are
momentum independent \cite{Rosiek:1989rs}.
Thus, a squark loop connecting gluons, $Z$ bosons, and $\phi$ bosons vanishes
upon adding its counterpart with the loop-momentum flows reversed if the total 
number of gluons and $Z$ bosons is odd.

As in Refs.~\cite{BarrientosBendezu:1999vd,BarrientosBendezu:2000tu}, we
express the quark and squark loop contributions in terms of helicity
amplitudes.
We label the helicity states of the two gluons and the $Z$ boson in the
partonic c.m.\ frame by $\lambda_a=\pm1/2$, $\lambda_b=\pm1/2$, and
$\lambda_Z=0,\pm1$.
We first consider $gg\to Z\phi$ with $\phi=h^0,H^0$.
The helicity amplitudes of the quark triangle contribution read
\begin{eqnarray}
{\mathcal M}_{\lambda_a\lambda_b0}^\triangle&=&
-2\sqrt{\frac{\lambda}{z}}(\lambda_a+\lambda_b)\sum_q
\left[\frac{z-s}{z}\ a_{Zqq}g_{\phi ZZ}{\mathcal P}_Z(s)
\left(F_\triangle\left(s,m_q^2\right)+2\right)
\right.\nonumber\\
&&{}-\left.\frac{s}{m_q}g_{A^0qq}g_{\phi A^0Z}{\mathcal P}_{A^0}(s)
F_\triangle\left(s,m_q^2\right)\right].
\label{eq:trih}
\end{eqnarray}
The quark triangle form factor, $F_\triangle$, is given in Eq.~(\ref{eq:fht}).
As for the quark box contribution, all twelve helicity combinations
contribute.
Due to Bose symmetry, they are related by
\begin{eqnarray}
{\mathcal M}_{\lambda_a\lambda_b\lambda_Z}^\Box(t,u)
&=&(-1)^{\lambda_Z}{\mathcal M}_{\lambda_b\lambda_a\lambda_Z}^\Box(u,t),
\nonumber\\
{\mathcal M}_{\lambda_a\lambda_b\lambda_Z}^\Box(t,u)
&=&{\mathcal M}_{-\lambda_a-\lambda_b-\lambda_Z}^\Box(t,u).
\label{eq:bos}
\end{eqnarray}
Keeping $\lambda_Z=\pm1$ generic, we thus only need to specify four
expressions.
These read
\begin{eqnarray}
{\mathcal M}_{++0}^\Box&=&
\frac{8}{\sqrt{z\lambda}}\sum_qg_{\phi qq}a_{Zqq}m_q
\left[F_{++}^0+(t\leftrightarrow u)\right],
\nonumber\\
{\mathcal M}_{+-0}^\Box&=&
\frac{8}{\sqrt{z\lambda}}\sum_qg_{\phi qq}a_{Zqq}m_q
\left[F_{+-}^0-(t\leftrightarrow u)\right],
\nonumber\\
{\mathcal M}_{++\lambda_Z}^\Box&=&
-4\sqrt{\frac{2N}{s}}\sum_qg_{\phi qq}a_{Zqq}m_q
\left[F_{++}^1-(t\leftrightarrow u)\right],
\nonumber\\
{\mathcal M}_{+-\lambda_Z}^\Box&=&
-4\sqrt{\frac{2N}{s}}\sum_qg_{\phi qq}a_{Zqq}m_q
\left[F_{+-}^1+(t\leftrightarrow u,\lambda_Z\leftrightarrow -\lambda_Z)\right].
\label{eq:boxh}
\end{eqnarray}
The quark box form factors, $F_{\lambda_a\lambda_b}^{|\lambda_Z|}$, are listed
in Eq.~(\ref{eq:fhb}).
For the reasons explained above, we have
$\tilde{\mathcal M}_{\lambda_a\lambda_b\lambda_Z}^\triangle
=\tilde{\mathcal M}_{\lambda_a\lambda_b\lambda_Z}^\Box=0$
for the squark-induced helicity amplitudes.

We now turn to $gg\to ZA^0$.
The helicity amplitudes of the quark and squark triangle contributions read
\begin{eqnarray}
{\mathcal M}_{\lambda_a\lambda_b0}^\triangle&=&
-8\sqrt{\frac{\lambda}{z}}(1+\lambda_a\lambda_b)\sum_qm_q\left(
g_{h^0A^0Z}g_{h^0qq}{\mathcal P}_{h^0}(s)+g_{H^0A^0Z}g_{H^0qq}
{\mathcal P}_{H^0}(s)
\right)F_\triangle\left(s,m_q^2\right),
\nonumber\\
\label{eq:qtria}\\
\tilde{\mathcal M}_{\lambda_a\lambda_b0}^\triangle&=&
2\sqrt{\frac{\lambda}{z}}(1+\lambda_a\lambda_b)\sum_{\tilde q_i}
\left(g_{h^0A^0Z}g_{h^0\tilde q_i\tilde q_i}{\mathcal P}_{h^0}(s)
+g_{H^0A^0Z}g_{H^0\tilde q_i\tilde q_i}{\mathcal P}_{H^0}(s)\right)
\tilde F_\triangle\left(s,m_{\tilde q_i}^2\right),
\nonumber\\
\label{eq:stria}
\end{eqnarray}
respectively.
The quark and squark triangle form factors, $F_\triangle$ and
$\tilde F_\triangle$, may be found in Eq.~(\ref{eq:fat}).
Again, the helicity amplitudes of the quark box contribution satisfy the Bose
symmetry relations of Eq.~(\ref{eq:bos}).
We find
\begin{eqnarray}
{\mathcal M}_{++0}^\Box&=&
-\frac{8}{\sqrt{z\lambda}}\sum_qg_{A^0qq}a_{Zqq}m_q
\left[F_{++}^0+(t\leftrightarrow u)\right],
\nonumber\\
{\mathcal M}_{+-0}^\Box&=&
-\frac{8}{\sqrt{z\lambda}}\sum_qg_{A^0qq}a_{Zqq}m_q
\left[F_{+-}^0+(t\leftrightarrow u)\right],
\nonumber\\
{\mathcal M}_{++\lambda_Z}^\Box&=&
-4\sqrt{\frac{2N}{s}}\sum_qg_{A^0qq}a_{Zqq}m_q
\left[F_{++}^1-(t\leftrightarrow u)\right],
\nonumber\\
{\mathcal M}_{+-\lambda_Z}^\Box&=&
-4\sqrt{\frac{2N}{s}}\sum_qg_{A^0qq}a_{Zqq}m_q
\left[F_{+-}^1-(t\leftrightarrow u,\lambda_Z\to-\lambda_Z)\right].
\end{eqnarray}
The quark box form factors, $F_{\lambda_a\lambda_b}^{|\lambda_Z|}$, are
presented in Eq.~(\ref{eq:fab}).
We recall that $\tilde{\mathcal M}_{\lambda_a\lambda_b\lambda_Z}^\Box=0$.

The differential cross section of $gg\to Z\phi$ is then given by
\begin{equation}
\frac{d\sigma}{dt}(gg\to Z\phi)=\frac{\alpha_s^2(\mu_r)G_F^2m_W^4}
{256(4\pi)^3s^2}\sum_{\lambda_a,\lambda_b,\lambda_Z}\left|
{\mathcal M}_{\lambda_a\lambda_b\lambda_Z}^\triangle
+{\mathcal M}_{\lambda_a\lambda_b\lambda_Z}^\Box
+\tilde{\mathcal M}_{\lambda_a\lambda_b\lambda_Z}^\triangle\right|^2,
\label{eq:xs}
\end{equation}
where $\alpha_s(\mu_r)$ is the strong-coupling constant at renormalization
scale $\mu_r$.
Due to Bose symmetry, the right-hand side of Eq.~(\ref{eq:xs}) is symmetric in
$t$ and $u$.

The differential cross section of $gg\to ZH$ in the SM is obtained from
Eqs.~(\ref{eq:trih})--(\ref{eq:boxh}) and (\ref{eq:xs}), with $\phi=h^0$, by
replacing $h^0\to H$, adjusting the $h^0ZZ$ and $h^0qq$ couplings, and
discarding the contribution due to $A^0$-boson exchange.
In this way, we recover the result of Ref.~\cite{Kniehl:1990iva}, which is
expressed in terms of Lorentz-invariant form factors rather than helicity
amplitudes.

The kinematics of the inclusive reaction $AB\to Z\phi+X$, where $A$ and $B$
are colliding hadrons, is described in Sec.~II of
Ref.~\cite{BarrientosBendezu:1998gd}.
Its double-differential cross section $d^2\sigma/dy\,dp_T$, where $y$ and
$p_T$ are the rapidity and transverse momentum of the $Z$ boson in the c.m.\
system of the hadronic collision, may be evaluated from Eq.~(2.1) of
Ref.~\cite{BarrientosBendezu:1998gd}.

\section{\label{sec:three}Phenomenological Implications}

We are now in a position to explore the phenomenological implications of our
results.
The SM input parameters for our numerical analysis are taken to be
$G_F=1.16637\times10^{-5}$~GeV$^{-2}$, $m_W=80.399$~GeV, $m_Z=91.1876$~GeV,
$m_t=172.0$~GeV , and $\overline{m}_b(\overline{m}_b)=4.19$~GeV
\cite{Nakamura:2010zzi}.
We adopt the LO proton PDF set CTEQ6L1 \cite{Pumplin:2002vw}.
We evaluate $\alpha_s(\mu_r)$ and $m_b(\mu_r)$ from the LO formulas, which may
be found, {\it e.g.}, in Eqs.~(23) and (24) of Ref.~\cite{Kniehl:1994dz},
respectively, with $n_f=5$ quark flavors and asymptotic scale parameter
$\Lambda_{\mathrm QCD}^{(5)}=165$~MeV \cite{Pumplin:2002vw}.
We identify the renormalization and factorization scales with the $Z\phi$
invariant mass $\sqrt s$.
We vary $\tan\beta$ and $m_{A^0}$ in the ranges
$3<\tan\beta<32\approx m_t/m_b$ and 180~GeV${}<m_{A^0}<1$~TeV, respectively.
As for the GUT parameters, we choose $m_{1/2}=150$~GeV, $A=0$, and $\mu<0$, 
and tune $m_0$ so as to be consistent with the desired value of $m_{A^0}$.
All other MSSM parameters are then determined according to the SUGRA-inspired
scenario as implemented in the program package SUSPECT \cite{Djouadi:2002ze}.
We do not impose the unification of the $\tau$-lepton and $b$-quark Yukawa
couplings at the GUT scale, which would just constrain the allowed $\tan\beta$
range without any visible effect on the results for these values of
$\tan\beta$.
We exclude solutions which do not comply with the present experimental lower
mass bounds of the sfermions, charginos, neutralinos, and Higgs bosons
\cite{Nakamura:2010zzi}.

\begin{figure}[ht]
\begin{center}
\begin{tabular}{c}
\parbox{15.5cm}{\epsfig{figure=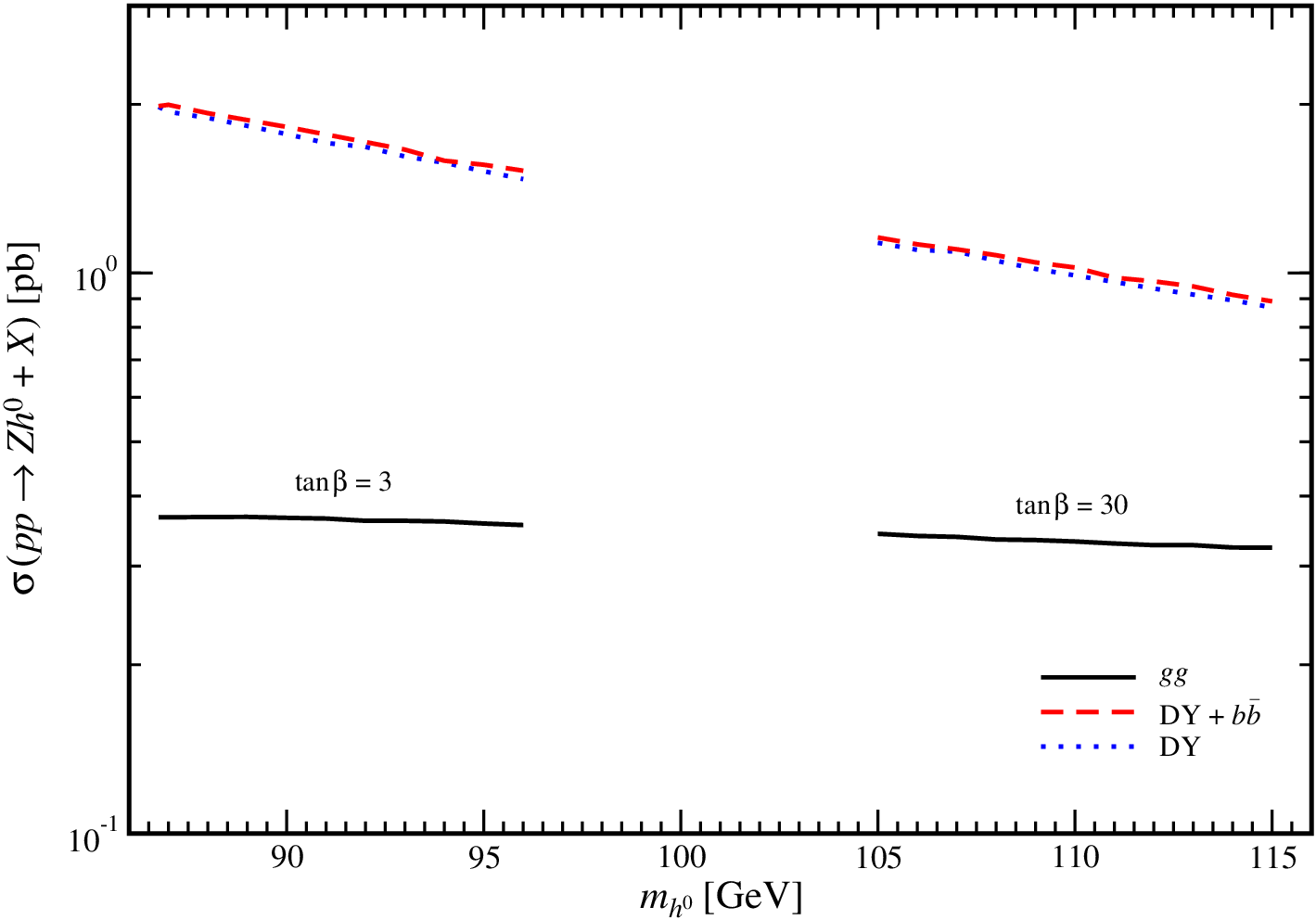,width=15.5cm}} \\
\\
\\
(a)
\end{tabular}
\caption{Total cross sections $\sigma$ (in pb) of $pp\to Zh^0+X$ via
$q\bar q$ annihilation (dashed lines) and $gg$ fusion (solid lines) at the LHC
(a) as functions of $m_{h^0}$ for $\tan\beta=3$ and 30; and (b) as functions
of $\tan\beta$ for $m_{A^0}=300$~GeV and $600$~GeV.
For comparison, also the Drell-Yan contribution to $q\bar q$ annihilation
(dotted lines) is shown.}
\label{fig:Zh}
\end{center}
\end{figure}

We now study the fully integrated cross sections of $pp\to Z\phi+X$ at the
LHC, with c.m.\ energy $\sqrt S=14$~TeV.
Figures~\ref{fig:Zh}--\ref{fig:ZA} refer to the cases $\phi=h^0,H^0,A^0$,
respectively.
In part (a) of each figure, the $m_\phi$ dependence is studied for
$\tan\beta=3$ and 30 while, in part~(b), the $\tan\beta$ dependence is studied
for $m_{A^0}=300$ and 600~GeV.
We note that the SUGRA-inspired MSSM with our choice of input parameters does
not permit $\tan\beta$ and $m_{A^0}$ to be simultaneously small, due to the
experimental lower bound on $m_{h^0}$ \cite{Nakamura:2010zzi}.
This explains why the curves for $\tan\beta=3$ in 
Figs.~\ref{fig:Zh}--\ref{fig:ZA}(a) only start at $m_{A^0}\approx280$~GeV,
while those for $\tan\beta=30$ already start at $m_{A^0}\approx180$~GeV.

In Figs.~\ref{fig:Zh} and \ref{fig:ZH}, which refer to $\phi=h^0,H^0$, 
respectively, the total $q\bar q$-annihilation contributions (dashed lines),
corresponding to the coherent superposition of Drell-Yan and Yukawa-enhanced
amplitudes, and the $gg$-fusion contributions (solid lines), which arise only
from quark loops, are presented separately.
For a comparison with future experimental data, they should be added.
For comparison, also the pure Drell-Yan contributions (dotted lines) are shown.
As for $\phi=h^0$, we observe from Fig.~\ref{fig:Zh} that the contribution due
to $q\bar q$ annihilation is almost exhausted by the Drell-Yan process and
greatly exceeds the one due to $gg$ fusion, by a factor of 3--5.
The $q\bar q$-annihilation contribution falls off by a factor of two as
$m_{h^0}$ runs from 82~GeV to 115~GeV and feebly depends on $\tan\beta$, except
for the appreciable rise towards the lower edge of the considered $\tan\beta$
range.
The $gg$-fusion contribution feebly depends on $m_{h^0}$, $m_{A^0}$, and
$\tan\beta$.
The situation is very different for $\phi=H^0$, as is obvious from
Fig.~\ref{fig:ZH}.
Here, $b\overline{b}$ annihilation is generally far more important than the
Drell-Yan process, except for $m_{A^0}=300$~GeV and $\tan\beta=3$, where the
latter gets close.
The contribution due to $b\overline{b}$ annihilation monotonically increases
with $\tan\beta$, while the one due to the Drell-Yan process decreases.
Furthermore, $gg$ fusion competes with $q\bar q$ annihilation and even
dominates for $\tan\beta\alt7$.

As for $\phi=A^0$, the $b\bar b$-annihilation contribution (dashed lines) and
the total $gg$-fusion contribution (solid lines), corresponding to the coherent
superposition of quark and squark loop amplitudes, are presented separately in
Fig.~\ref{fig:ZA}.
For comparison, also the $gg$-fusion contribution due to quark loops only
(dotted  lines) is shown.
As in the case of $\phi=H^0$, $gg$ fusion competes with $b\bar b$ annihilation
and even dominates for $\tan\beta\alt7$.
Again, the $b\overline{b}$-annihilation contribution monotonically increases
with $\tan\beta$.
The bulk of the $gg$-fusion contribution is due to the quark loops, especially
at low values of $m_{A^0}$.

Finally, we compare our results with the literature.
As already mentioned in Sec.~\ref{sec:two}, we recover the well-known SM
result \cite{Kniehl:1990iva}, for $\phi=H$, by taking the SM limit of our
results for $\phi=h^0$ in Eqs.~(\ref{eq:trih}) and (\ref{eq:boxh}).
The contribution due to $A^0$-boson exchange in Eq.~(\ref{eq:trih}), which is
not probed in the SM limit, agrees with the analogous contribution to
$gg\to W^-H^+$ given in Eq.~(1) of Ref.~\cite{BarrientosBendezu:1999vd} after
appropriately adjusting the masses and couplings.
On the other hand, the residual terms in the latter equation, which arise from
the exchanges of $h^0$ and $H^0$ bosons, coincide with Eq.~(\ref{eq:qtria})
after substituting the appropriate masses and couplings.
Similarly, by adjusting masses and couplings in Eq.~(\ref{eq:stria}), we
reproduce Eq.~(2.3) in Ref.~\cite{BarrientosBendezu:2000tu}, which gives the
squark triangle contribution to $gg\to W^-H^+$.
In Ref.~\cite{Kao:1991xg}, numerical results for the cross section of
$pp\to ZA^0$ via quark-loop-mediated $gg$ fusion were presented.
Adopting the input parameters and proton PDF set specified in that reference,
we nicely reproduce the separate contributions due triangle and box diagrams
shown in Fig.~4 therein, while we fail to agree with their superposition.
Furthermore, we find reasonable agreement with the cross section of
$pp\to ZA^0+X$ via $gg$ fusion represented graphically for different scenarios
in Figs.~6 and 7 of Ref.~\cite{Li:2005qna} adopting the respective inputs from
there.

\section{\label{sec:four}Conclusions}

We analytically calculated the cross sections of the partonic subprocesses
$q\bar q\to Z\phi$ and $gg\to Z\phi$, where $\phi=h^0,H^0,A^0$, to LO in the
MSSM.
We included the Drell-Yan and Yukawa-enhanced contributions to $q\bar q$ 
annihilation (see Fig.~\ref{fig:tree}) and the quark and squark loop
contributions to $gg$ fusion (see Fig.~\ref{fig:loop}).
We presented these results as helicity amplitudes expressed in terms of
standard scalar one-loop integrals.

We then quantitatively investigated the inclusive cross sections of
$pp\to Z\phi+X$ at the LHC with $\sqrt{S}=14$~GeV adopting a favorable
SUGRA-inspired MSSM scenario, varying the input parameters $m_{A^0}$ and
$\tan\beta$.
Our results are presented in Figs.~\ref{fig:Zh}--\ref{fig:ZA}.
The total cross section for $\phi=h^0$ is typically of order 1~pb, while those
for $\phi=H^0,A^0$ are of order 100~fb (10~fb) for $m_{A^0}=300$~GeV (600~GeV).
Assuming design luminosity, $L=10^{34}$~cm${}^{-2}$s${}^{-1}$, a cross section
of 1~pb corresponds to $10^5$ events per year and experiment at the LHC (see
Table~I of Ref.~\cite{Kniehl:2002wd}).

\section*{Acknowledgments}

We thank A.~A.~Barrientos Bendezu and R.~Ziegler for their collaboration at the
initial stage of this work.
The work of B.A.K. was supported in part by the German Federal Ministry for
Education and Research BMBF through Grant No.\ 05~HT6GUA, by the German
Research Foundation DFG through the Collaborative Research Centre No.~676
{\it Particles, Strings and the Early Universe---The Structure of Matter and
Space Time}, and by the Helmholtz Association HGF through the Helmholtz
Alliance Ha~101 {\it Physics at the Terascale}.
The work of C.P.P. was supported in part by the German Academic Exchange
Service (DAAD) Reinvitation Programme under Reference Code A/07/02820 and
by the Office of the Vice President for Academic Affairs of the University of
the Philippines.

\def\theequation{\Alph{section}.\arabic{equation}}
\begin{appendix}
\setcounter{equation}{0}

\section{\label{sec:a}Feynman rules}

In this appendix, we collect the Feynman rules used in this paper.
The Feynman rules for the $Zq\overline{q}$ vertices are
$ig\gamma^\mu(v_{Zqq}-a_{Zqq}\gamma_5)$, with $g=e/s_w$, $e$ being the proton
charge, $s_w^2=1-c_w^2$, and
\begin{equation}
v_{Zqq}=-\frac{I_q-2s_w^2Q_q}{2c_w},\qquad
a_{Zqq}=-\frac{I_q}{2c_w},
\end{equation}
where $I_q=\pm1/2$ and $Q_q=2/3,-1/3$ are the weak hypercharge and electric
charge of quark $q$, respectively.
The Feynman rules for the $\phi q\overline{q}$ ($\phi=h^0,H^0$) and
$A^0q\overline{q}$ vertices are $igg_{\phi qq}$ and $gg_{A^0qq}\gamma_5$,
respectively, with
\begin{eqnarray}
g_{h^0tt}=-\frac{m_t\cos\alpha}{2m_W\sin\beta},\qquad
g_{H^0tt}=-\frac{m_t\sin\alpha}{2m_W\sin\beta},\qquad
g_{A^0tt}=-\frac{m_t\cot\beta}{2m_W},
\nonumber\\
g_{h^0bb}=\frac{m_b\sin\alpha}{2m_W\cos\beta},\qquad
g_{H^0bb}=-\frac{m_b\cos\alpha}{2m_W\cos\beta},\qquad
g_{A^0bb}=-\frac{m_b\tan\beta}{2m_W},
\end{eqnarray}
where $\alpha$ is the mixing angle that rotates the weak $CP$-even Higgs
eigenstates into the mass eigenstates $h^0$ and $H^0$.
The Feynman rules for the $\phi ZZ$ vertices are $igg_{\phi ZZ}g^{\mu\nu}$,
with
\begin{equation}
g_{h^0ZZ}=-\frac{m_Z}{c_w}\sin(\alpha-\beta),\qquad
g_{H^0ZZ}=\frac{m_Z}{c_w}\cos(\alpha-\beta).
\label{eq:hzz}
\end{equation}
The Feynman rules for the $\phi A^0Z$ vertices are
$gg_{\phi A^0Z}(p+p^\prime)^\mu$, where $p$ is the incoming four-momentum of
the $\phi$ boson, $p^\prime$ is the outgoing four-momentum of the $A^0$ boson, 
and
\begin{equation}
g_{h^0A^0Z}=\frac{\cos(\alpha-\beta)}{2c_w},\qquad
g_{H^0A^0Z}=\frac{\sin(\alpha-\beta)}{2c_w}.
\end{equation}
The Feynman rules for the $\phi\tilde{q}_i\tilde{q}_j$ vertices are
$igg_{\phi\tilde{q}_i\tilde{q}_j}$, with
\begin{eqnarray}
\lefteqn{\left(\begin{array}{cc}
g_{h^0\tilde{t}_1\tilde{t}_1} & g_{h^0\tilde{t}_1\tilde{t}_2} \\
g_{h^0\tilde{t}_2\tilde{t}_1} & g_{h^0\tilde{t}_2\tilde{t}_2} \\
\end{array}\right)}
\nonumber\\
&=&{\cal M}^{\tilde{t}}\left(\begin{array}{cc}
\frac{m_Z\sin(\alpha+\beta)\left(I^3_t-s_w^2Q_t\right)}{c_w}
-\frac{m_t^2\cos\alpha}{m_W\sin\beta} 
& -\frac{m_t\left(\mu\sin\alpha+A_t\cos\alpha\right)}{2m_W\sin\beta} \\
-\frac{m_t\left(\mu\sin\alpha+A_t\cos\alpha\right)}{2m_W\sin\beta} 
& \frac{m_Z\sin(\alpha+\beta)s_w^2Q_t}{c_w}-\frac{m_t^2\cos\alpha}
{m_W\sin\beta} \\
\end{array}\right)\left({\cal M}^{\tilde{t}}\right)^T,
\nonumber\\
\lefteqn{\left(\begin{array}{cc}
g_{h^0\tilde{b}_1\tilde{b}_1} & g_{h^0\tilde{b}_1\tilde{b}_2} \\
g_{h^0\tilde{b}_2\tilde{b}_1} & g_{h^0\tilde{b}_2\tilde{b}_2} \\
\end{array}\right)}
\nonumber\\
&=& {\cal M}^{\tilde{b}}\left(\begin{array}{cc}
\frac{m_Z\sin(\alpha+\beta)\left(I^3_b-s_w^2Q_b\right)}{c_w}
+\frac{m_b^2\sin\alpha}{m_W\cos\beta} 
& \frac{m_b\left(\mu\cos\alpha+A_b\sin\alpha\right)}{2m_W\cos\beta} \\
\frac{m_b\left(\mu\cos\alpha+A_b\sin\alpha\right)}{2m_W\cos\beta}
& \frac{m_Z\sin(\alpha+\beta)s_w^2Q_b}{c_w}
+\frac{m_b^2\sin\alpha}{m_W\cos\beta} \\
\end{array}\right)\left({\cal M}^{\tilde{b}}\right)^T,
\nonumber\\
\lefteqn{\left(\begin{array}{cc}
g_{H^0\tilde{t}_1\tilde{t}_1} & g_{H^0\tilde{t}_1\tilde{t}_2} \\
g_{H^0\tilde{t}_2\tilde{t}_1} & g_{H^0\tilde{t}_2\tilde{t}_2} \\
\end{array}\right)}
\nonumber\\
&=& {\cal M}^{\tilde{t}}\left(\begin{array}{cc}
-\frac{m_Z\cos(\alpha+\beta)\left(I^3_t-s_w^2Q_t\right)}{c_w}
-\frac{m_t^2\sin\alpha}{m_W\sin\beta} &
\frac{m_t\left(\mu\cos\alpha-A_t\sin\alpha\right)}{2m_W\sin\beta} \\
\frac{m_t\left(\mu\cos\alpha-A_t\sin\alpha\right)}{2m_W\sin\beta} 
& -\frac{m_Z\cos(\alpha+\beta)s_w^2Q_t}{c_w}
-\frac{m_t^2\sin\alpha}{m_W\sin\beta} \\
\end{array}\right)\left({\cal M}^{\tilde{t}}\right)^T,
\nonumber\\
\lefteqn{\left(\begin{array}{cc}
g_{H^0\tilde{b}_1\tilde{b}_1} & g_{H^0\tilde{b}_1\tilde{b}_2} \\
g_{H^0\tilde{b}_2\tilde{b}_1} & g_{H^0\tilde{b}_2\tilde{b}_2} \\
\end{array}\right)}
\nonumber\\
&=& {\cal M}^{\tilde{b}}\left(\begin{array}{cc}
-\frac{m_Z\cos(\alpha+\beta)\left(I^3_b-s_w^2Q_b\right)}{c_w}
-\frac{m_b^2\cos\alpha}{m_W\cos\beta} 
& \frac{m_b\left(\mu\sin\alpha-A_b\cos\alpha\right)}{2m_W\cos\beta} \\
\frac{m_b\left(\mu\sin\alpha-A_b\cos\alpha\right)}{2m_W\cos\beta} 
& -\frac{m_Z\cos(\alpha+\beta)s_w^2Q_b}{c_w}
-\frac{m_b^2\cos\alpha}{m_W\cos\beta} \\
\end{array}\right)\left({\cal M}^{\tilde{b}}\right)^T,
\end{eqnarray}
where
\begin{equation}
{\cal M}^{\tilde{q}}=\left(\begin{array}{cc}
\cos\theta_{\tilde{q}}\ & \sin\theta_{\tilde{q}} \\
-\sin\theta_{\tilde{q}}\ & \cos\theta_{\tilde{q}} \\
\end{array}\right)
\end{equation}
are the squark mixing matrices, with $\theta_{\tilde{q}}$ being the squark
mixing angles.

\section{\label{sec:b}Quark and squark loop form factors}

In this appendix, we express the quark and squark triangle and box form
factors, $F_\triangle$, $\tilde F_\triangle$, and $F_{\lambda_a\lambda_b}^{|\lambda_Z|}$,
for $\phi=h^0,H^0$ and $\phi=A^0$, in terms of the standard scalar three- and
four-point functions, which we abbreviate as
$C_{ijk}^{ab}(c)=C_0\left(a,b,c,m_i^2,m_j^2,m_k^2\right)$ and
$D_{ijkl}^{abcd}(e,f)=D_0\left(a,b,c,d,e,f,m_i^2,m_j^2,m_k^2,m_l^2\right)$,
respectively.
The definitions of the latter may be found in Eq.~(5) of
Ref.~\cite{BarrientosBendezu:1999vd}.

The quark triangle form factor for $\phi=h^0,H^0$ reads
\begin{equation}
F_\triangle\left(s,m_q^2\right)=4m_q^2C_{qqq}^{00}(s).
\label{eq:fht}
\end{equation}

The quark box form factors for $\phi=h^0,H^0$ read
\begin{eqnarray}
F_{++}^0&=&2s(t+u)C^{00}_{qqq}(s)
+2\left(t+u+\frac{\lambda}{s}\right)
\left[(t-z)C^{z0}_{qqq}(t)+(t-h)C^{h0}_{qqq}(t)\right]
\nonumber\\
&&{}-\left[N\left(t+u+\frac{\lambda}{s}\right)+2m_q^2\lambda\right]
D^{h0z0}_{qqqq}(t,u)-4\left(szh+m_q^2\lambda\right)D^{hz00}_{qqqq}(s,t),
\nonumber\\
F_{+-}^0&=&\frac{(h-z-s)}{N}(t-u)\left[s(t+u)C^{00}_{qqq}(s)
-\lambda C^{hz}_{qqq}(s)-2m_q^2ND^{h0z0}_{qqqq}(t,u)\right]
\nonumber\\
&&{}+2(t+u)(t-z)\left[1+\frac{t(t-u)(h-z-s)}{N(t+u)}\right]C^{z0}_{qqq}(t)
\nonumber\\
&&{}+\frac{2(t-h)}{N}\left[z(u^2-t^2-\lambda)+(t+u)(t^2-zh)\right]C^{h0}_{qqq}(t)
\nonumber\\
&&{}-(h-z-s)\left[2st\frac{t^2-zh}{N}+4m_q^2(t-u)\right]D^{hz00}_{qqqq}(s,t),
\nonumber\\
F_{++}^1&=&(t-u)\left[\frac{z-h-s}{\sqrt\lambda}-\lambda_Z\right]
\left[\frac{s}{N}C^{00}_{qqq}(s)-\frac{1}{2}D^{h0z0}_{qqqq}(t,u)-
\frac{s}{N}\left(t+\frac{2N}{t-u}\right)D^{hz00}_{qqqq}(s,t)\right]
\nonumber\\
&&{}+\frac{2(h-u)}{\sqrt\lambda N}\left(\lambda_Z\sqrt\lambda+t-u+\frac{2N}{h-u}\right)
\left[(h-t)C^{h0}_{qqq}(t)+(z-u)C^{z0}_{qqq}(u)\right],
\nonumber\\
F_{+-}^1&=&\frac{s}{N}\left(\frac{4s(t+u)}{\sqrt\lambda}+\sqrt\lambda -\lambda_Z(t-u)
\right)C^{00}_{qqq}(s)-\frac{2s}{N}\left(\sqrt\lambda 
+\lambda_Z(t-u)\right)C^{hz}_{qqq}(s)
\nonumber\\
&&{}-\frac{2(t-h)}{\sqrt\lambda N}\left(-s(u+3t)-2N+(u-t)(t-z)
+\lambda_Z(t-s-z)\sqrt\lambda\right)C^{h0}_{qqq}(t)
\nonumber\\
&&{}+\frac{2(u-z)}{\sqrt\lambda N}\left(3u(s-z)+th-2z(h-2u)-\lambda_Z(h-u)
\sqrt\lambda\right)C^{z0}_{qqq}(u)
\nonumber\\
&&{}+\frac{s}{\sqrt\lambda N}\left[t\left(\lambda+8zh-4ts
-2(t+u)(z+h)\right.\right.
\nonumber\\
&&{}+\left.\left.\lambda_Z(-2h+3t+u-2z)\sqrt\lambda \right)
-16m_q^2N\right]D^{hz00}_{qqqq}(s,t)
\nonumber\\
&&{}+\frac{1}{2}\left(-\sqrt\lambda-\frac{16m_q^2s}{\sqrt\lambda}+
\lambda_Z(t-u)\right)D^{h0z0}_{qqqq}(t,u).
\label{eq:fhb}
\end{eqnarray}

The quark and squark triangle form factors for $\phi=A^0$ read
\begin{eqnarray}
F_\triangle\left(s,m_q^2\right)&=&2+\left(4m_q^2-s\right)C^{00}_{qqq}(s),
\nonumber\\
\tilde F_\triangle\left(s,m_{\tilde q}^2\right)&=&
2+4m_{\tilde q_i}^2C^{00}_{\tilde q_i\tilde q_i\tilde q_i}(s).
\label{eq:fat}
\end{eqnarray}

The quark box form factors for $\phi=A^0$ read
\begin{eqnarray}
F^0_{++}&=&2s(t+u)C^{00}_{qqq}(s)
+2\left(t+u+\frac{\lambda}{s}\right)
\left[(t-z)C^{z0}_{qqq}(t)+(t-h)C^{h0}_{qqq}(t)\right]
\nonumber\\
&&{}-\left[N\left(t+u+\frac{\lambda}{s}\right)+2m_q^2\lambda\right]
D^{h0z0}_{qqqq}(t,u)-4\left(szh+m_q^2\lambda\right)D^{hz00}_{qqqq}(s,t),
\nonumber\\
F^0_{+-}&=&
-\left[2+\frac{(t-u)^2}{N}\right]\left[s(t+u)C^{00}_{qqq}(s)
-\lambda C^{hz}_{qqq}(s)\right]
\nonumber\\
&&{}-2\left[3t-u+\frac{t}{N}(t-u)^2\right]
\left[(t-z)C^{z0}_{qqq}(t)+(t-h)C^{h0}_{qqq}(t)\right]
\nonumber \\
&&{}-2\left(zN-m_q^2\lambda\right)D^{h0z0}_{qqqq}(t,u)
\nonumber\\
&&{}+2\left\{st\left[3t-u+\frac{t}{N}(t-u)^2\right]+2m_q^2\lambda\right\}
D^{hz00}_{qqqq}(s,t),
\nonumber\\
F^1_{++}&=&\left(\frac{z-h-s}{\sqrt\lambda}-\lambda_Z\right)
\left\{(t-u)\left(\frac{s}{N}C^{00}_{qqq}(s)-\frac{1}{2}D^{h0z0}_{qqqq}(t,u)
\right)\right.
\nonumber\\
&&{}-\left.s\left[2+\frac{t}{N}(t-u)\right]D^{hz00}_{qqqq}(s,t)\right\}
\nonumber\\
&&{}+2(t-z)\left\{\lambda_Z\frac{h-t}{N}+\frac{1}{\sqrt\lambda}
\left[2+\frac{(t-u)(t-h)}{N}\right]\right\}C^{z0}_{qqq}(t)
\nonumber\\
&&{}-2(t-h)\left\{\lambda_Z\frac{h-u}{N}+\frac{1}{\sqrt\lambda}
\left[2-\frac{(t-u)(u-h)}{N}\right]\right\}C^{h0}_{qqq}(t),
\nonumber\\
F^1_{+-}&=&\left(\lambda_Z-\frac{t-u}{\sqrt\lambda}\right)
\left[\frac{s}{N}(s-z+h)\left(C^{00}_{qqq}(s)-tD^{hz00}_{qqqq}(s,t)\right)
+\frac{s+z-h}{2}D^{h0z0}_{qqqq}(t,u)\right]
\nonumber\\
&&{}-2(t-z)\left\{\lambda_Z\frac{t-h}{N}-\frac{1}{\sqrt\lambda}
\left[2+\frac{(t-u)(t-h)}{N}\right]\right\}C^{z0}_{qqq}(t)
\nonumber\\
&&{}-2(t-h)\left\{\lambda_Z\frac{u-h}{N}+\frac{1}{\sqrt\lambda}
\left[2-\frac{(t-u)(u-h)}{N}\right]\right\}C^{h0}_{qqq}(t).
\label{eq:fab}
\end{eqnarray}

\end{appendix}

\newpage

\newpage

\newpage

\newpage
\begin{figure}[ht]
\begin{center}
\begin{tabular}{c}
\parbox{15.5cm}{\epsfig{figure=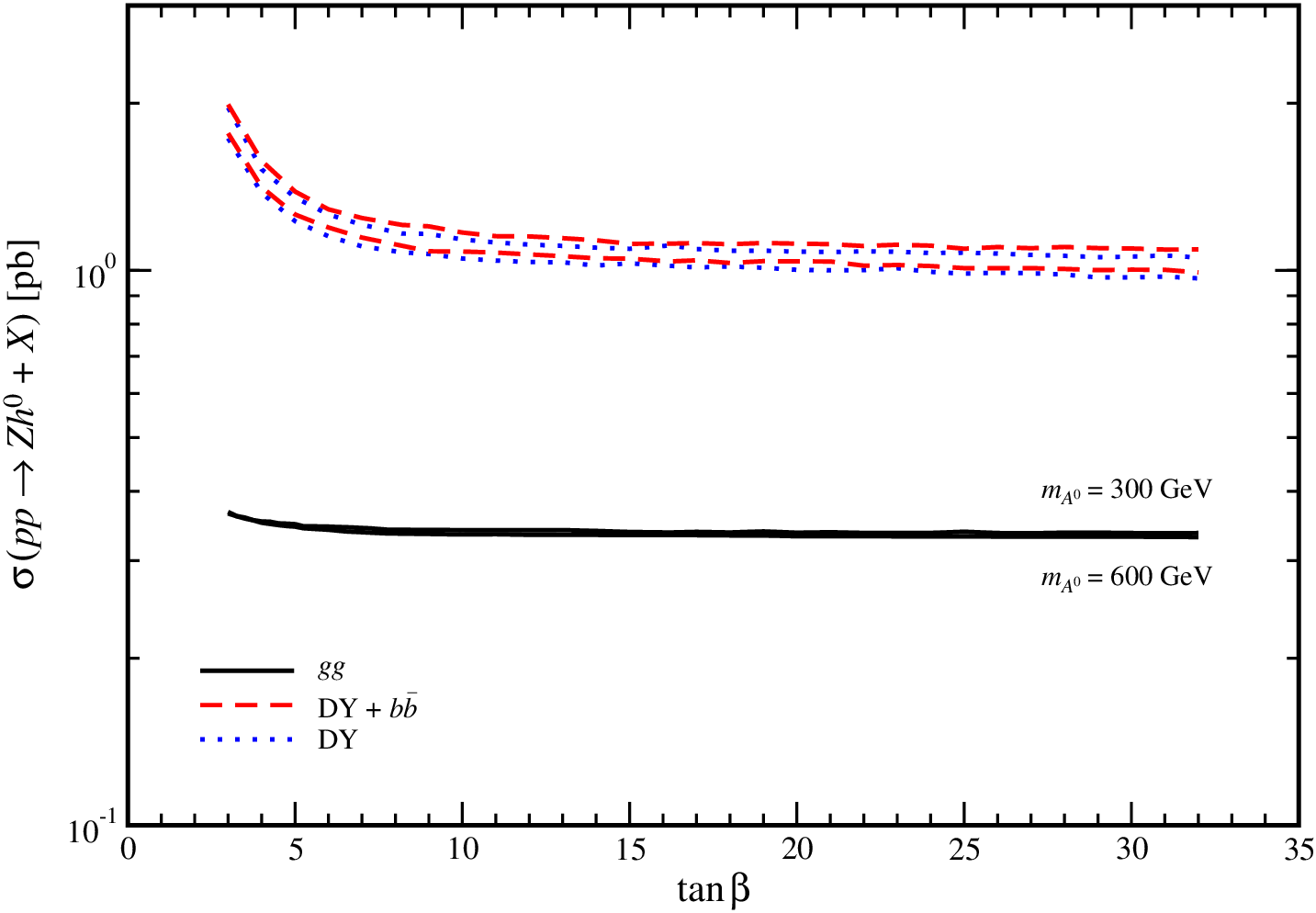,width=15.5cm}} \\
\\
\\
(b)
\end{tabular}
\vspace{1cm}
\\
Figure 3 (Continued).
\end{center}
\end{figure}

\newpage
\begin{figure}[ht]
\begin{center}
\begin{tabular}{c}
\parbox{15.5cm}{\epsfig{figure=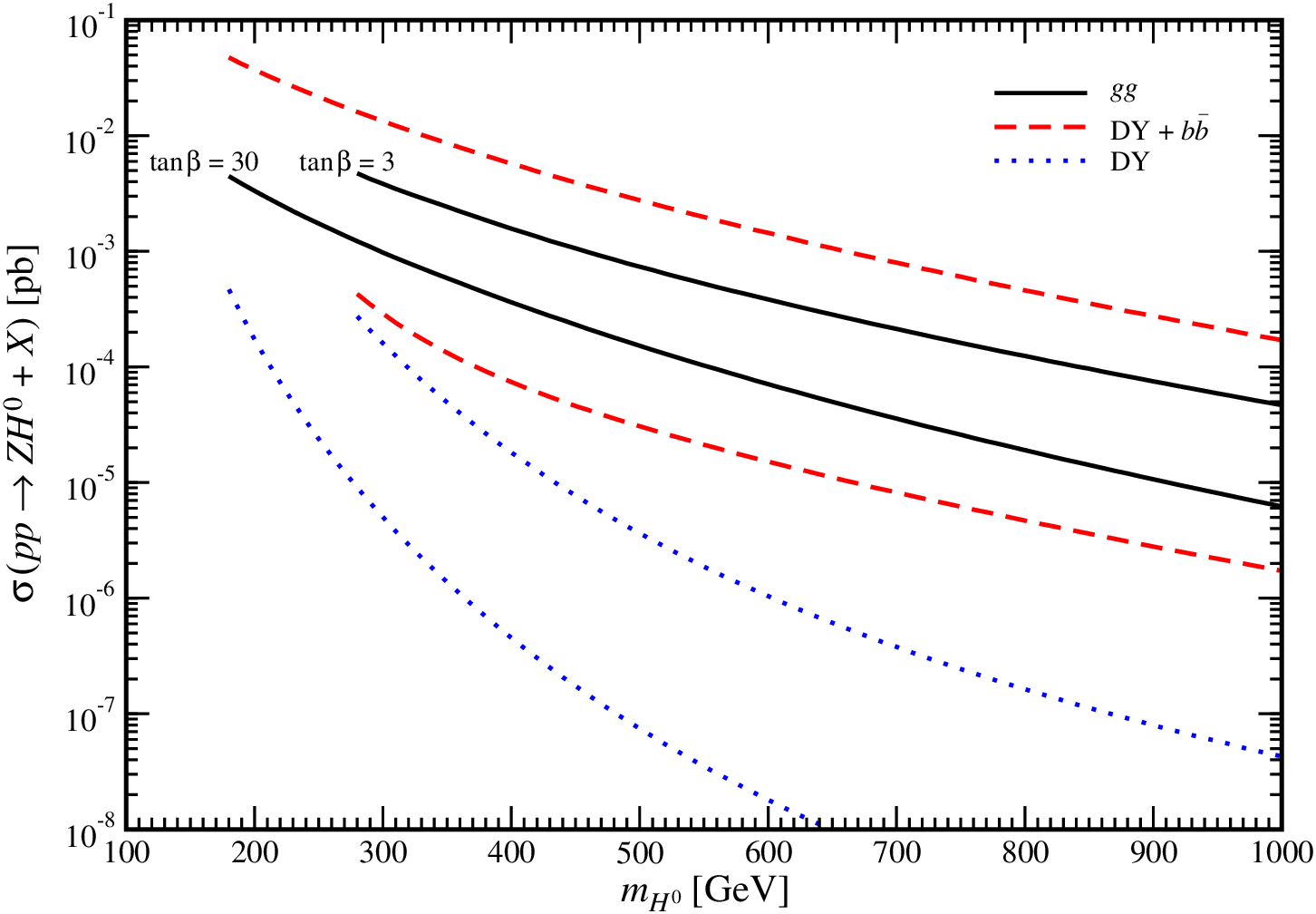,width=15.5cm}} \\
\\
\\
(a)
\end{tabular}
\caption{Total cross sections $\sigma$ (in pb) of $pp\to ZH^0+X$ via
$q\bar q$ annihilation (dashed lines) and $gg$ fusion (solid lines) at the LHC
(a) as functions of $m_{H^0}$ for $\tan\beta=3$ and 30; and (b) as functions
of $\tan\beta$ for $m_{A^0}=300$~GeV and $600$~GeV.
For comparison, also the Drell-Yan contribution to $q\bar q$ annihilation
(dotted lines) is shown.}
\label{fig:ZH}
\end{center}
\end{figure}

\newpage
\begin{figure}[ht]
\begin{center}
\begin{tabular}{c}
\parbox{15.5cm}{\epsfig{figure=tHH.eps,width=15.5cm}} \\
\\
\\
(b)
\end{tabular}
\vspace{1cm}
\\
Figure 4 (Continued).
\end{center}
\end{figure}

\newpage
\begin{figure}[ht]
\begin{center}
\begin{tabular}{c}
\parbox{15.5cm}{\epsfig{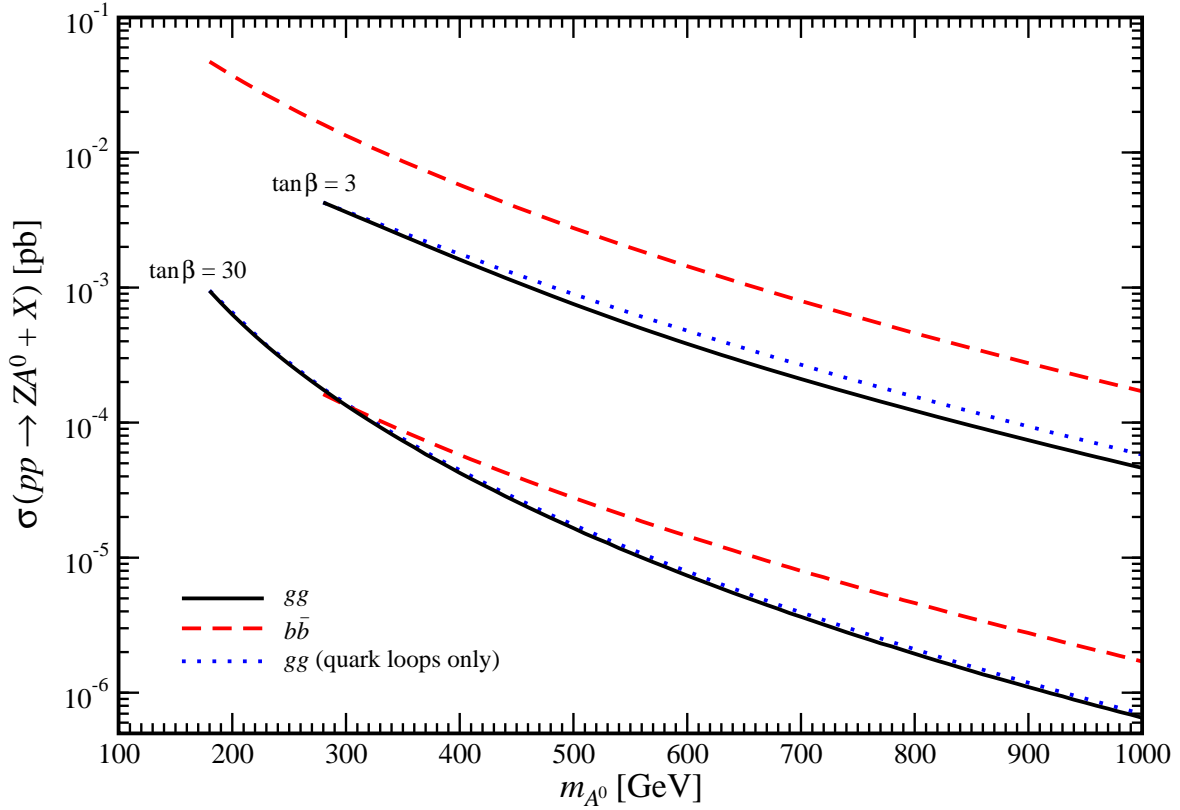}} \\
\\
\\
(a)
\end{tabular}
\caption{Total cross sections $\sigma$ (in pb) of $pp\to ZA^0+X$ via
$b\bar b$ annihilation (dashed lines) and $gg$ fusion (solid lines) at the LHC
(a) as functions of $m_{A^0}$ for $\tan\beta=3$ and 30; and (b) as functions
of $\tan\beta$ for $m_{A^0}=300$~GeV and $600$~GeV.
For comparison, also the quark loop contribution to $gg$ fusion (dotted lines)
is shown.}
\label{fig:ZA}
\end{center}
\end{figure}

\newpage
\begin{figure}[ht]
\begin{center}
\begin{tabular}{c}
\parbox{15.5cm}{\epsfig{figure=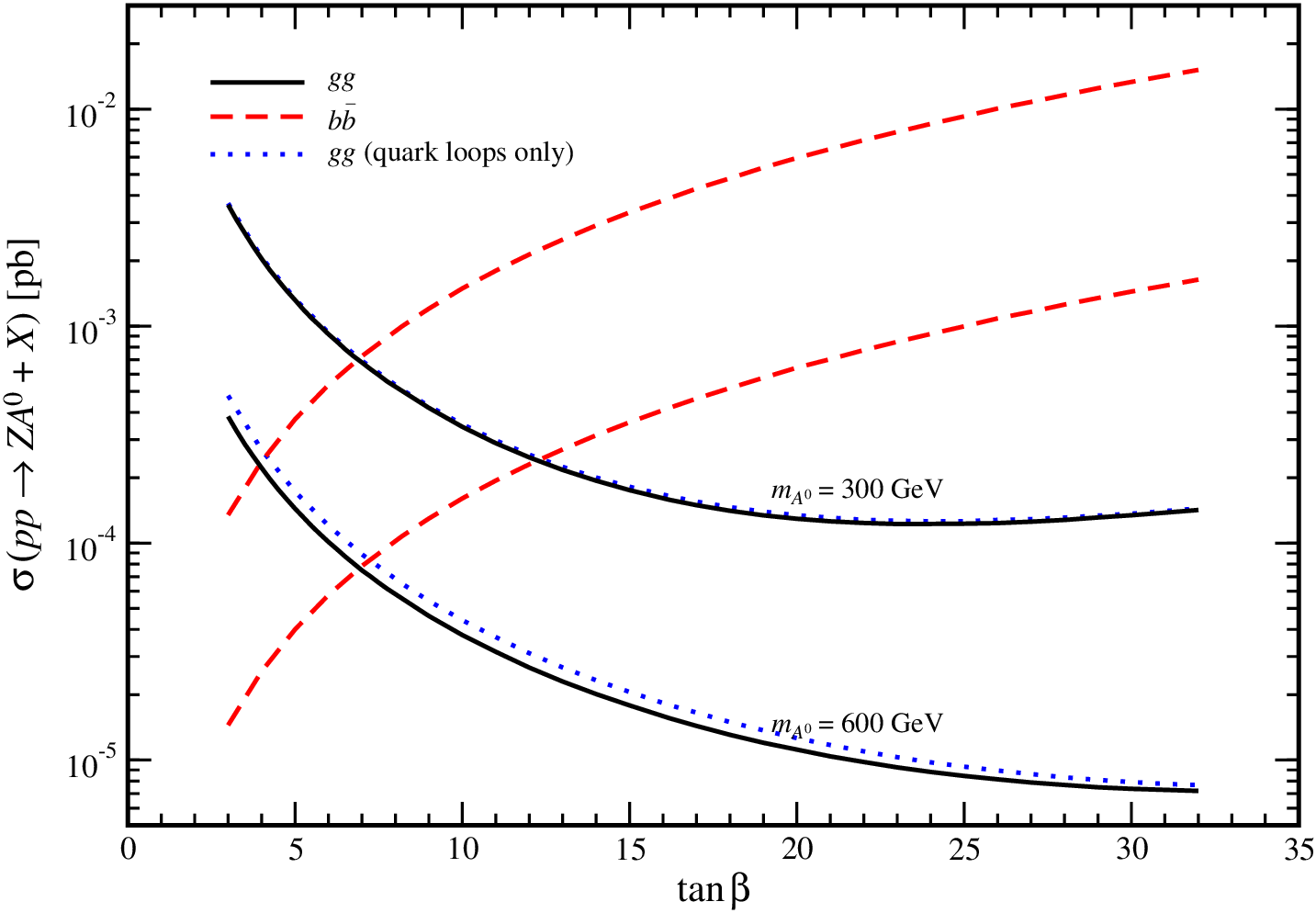,width=15.5cm}} \\
\\
\\
(b)
\end{tabular}
\vspace{1cm}
\\
Figure 5 (Continued).
\end{center}
\end{figure}

\end{document}